\documentclass[suppldata]{interact}

\usepackage[numbers,sort&compress,merge]{natbib}
\bibpunct[, ]{[}{]}{,}{n}{,}{,}

\usepackage{bbold}
\usepackage{cancel}
\usepackage{xfrac} 
\usepackage[nointegrals]{wasysym} 

\usepackage[pdftex,bookmarks,colorlinks,breaklinks,colorlinks=false, pdfborder={0 0 0}]{hyperref} 
\hypersetup{pdftitle={Scratching a 50-year itch with elongated rods},%
			pdfauthor={Ramon Creyghton and Bela Mulder},%
			pdfsubject={Nematic Liquid Crystals},%
			pdfkeywords={Theory, Liquid Crystals; Nematics; Density Functional Theory},%
			pdfcreator={}
}

\newcommand{\citeref}[1]{Ref.~\citenum{#1}}

\newcommand{\refeq}[1]{Eq.~\ref{#1}}

\newcommand{\qtens}{$\mathbf{Q}$-tensor}

\newcommand{\pr}{^{\prime}}

\newcommand{\dd}{\mathrm{d}}

\newcommand{\vr}{\mathbf{r}}

\newcommand{\ddr}{\dd\mathbf{r}}

\newcommand{\vp}{\mathbf{p}}

\newcommand{\vq}{\mathbf{q}}

\newcommand{\bss}{\boldsymbol}
\newcommand{\vom}{\boldsymbol{\hat{\omega}}}

\newcommand{\vomi}{\boldsymbol{\hat{\omega}}_{i}}

\newcommand{\ddom}{\dd\boldsymbol{\hat{\omega}}}

\newcommand{\omci}{\omega_{i}}
\newcommand{\omce}{\omega_{1}}
\newcommand{\omct}{\omega_{2}}
\newcommand{\omcd}{\omega_{3}}
\newcommand{\omcia}{\omega_{\alpha}}
\newcommand{\omcib}{\omega_{\beta}}
\newcommand{\omcig}{\omega_{\gamma}}
\newcommand{\omcid}{\omega_{\delta}}

\newcommand{\omcis}{\omega_{\sigma}}
\newcommand{\omcit}{\omega_{\tau}}

\newcommand{\vn}{\mathbf{\hat{n}}}
\newcommand{\vnci}{\vn_{i}}
\newcommand{\vnce}{\vn_{1}}
\newcommand{\vnct}{\vn_{2}}
\newcommand{\vncd}{\vn_{3}}

\newcommand{\ncia}{n_{\alpha}}
\newcommand{\ncib}{n_{\beta}}
\newcommand{\ncig}{n_{\gamma}}
\newcommand{\ncid}{n_{\delta}}

\newcommand{\ncis}{n_{\sigma}}
\newcommand{\ncit}{n_{\tau}}

\newcommand{\pda}{\partial_{\alpha}}
\newcommand{\pdb}{\partial_{\beta}}

\newcommand{\nn}[1]{^{\mbox{\tiny(#1)}}}
\newcommand{\nnz}{\nn{0}}

\newcommand{\nnt}{\nn{2}}

\newcommand{\kce}{k_{1}}
\newcommand{\kct}{k_{2}}
\newcommand{\kcd}{k_{3}}

\newcommand{\qce}{q_{1}}
\newcommand{\qct}{q_{2}}
\newcommand{\qcd}{q_{3}}
\newcommand{\va}{\bss\alpha}
\newcommand{\ace}{\alpha_{1}}
\newcommand{\act}{\alpha_{2}}
\newcommand{\acd}{\alpha_{3}}

\newcommand{\psirom}{\psi(\vr,\vom)}

\newcommand{\td}[1]{_{\mathrm{#1}}}

\newcommand{\thvol}{\mathcal{V}_{T}}

\newcommand{\F}{\mathcal{F}}
\newcommand{\Fid}{\F_{\mathrm{id}}}
\newcommand{\Z}{\mathcal{Z}}
\newcommand{\tQ}{\mathbf{Q}}
\newcommand{\tO}{\mathbf{O}}
\newcommand{\tA}{\mathbf{A}}

\newcommand{\tH}{\mathbf{H}}

\newcommand{\tM}{\mathbf{M}}
\newcommand{\eM}{\mathcal{M}}

\newcommand{\boh}{\mathcal{O}}

\newcommand{\vomvom}{\vom \otimes \vom}

\DeclareMathOperator{\Tr}{Tr}

\usepackage{xstring}
\newcommand{\ctp}[3]{%
  \IfEqCase{#1}{%
    {1}{\hspace{.3ex}{}_{\mbox{\tiny#2}}\hspace{-0.2ex}{\circ\hspace{-0.2ex}}_{\mbox{\tiny#3}}\hspace{.4ex}}%
    {2}{\hspace{.4ex}{}_{\mbox{\tiny#2}}\hspace{-0.6ex}\wr_{\mbox{\tiny#3}}\hspace{.1ex}}%
    {3}{\hspace{.5ex}{}_{\mbox{\tiny#2}}\hspace{-0.8ex}\vartriangle_{\mbox{\tiny#3}}\hspace{.1ex}}%
    {4}{\hspace{.5ex}{}_{\mbox{\tiny#2}}\hspace{-0.1ex}\square_{\mbox{\tiny#3}}\hspace{.5ex}}%
  }[\hspace{.5ex}{}_{\mbox{\tiny#2}}\hspace{-0.15ex}\overset{\mbox{\tiny#1}}{\Circle}_{\mbox{\tiny#3}}\hspace{.5ex}]%
}
\newcommand{\ctf}[1]{%
  \IfEqCase{#1}{%
    {1}{\hspace{-.1ex}\cdot\hspace{0ex}}%
    {2}{\hspace{.1ex}\colon\hspace{-.2ex}}%
    {3}{\hspace{.4ex}\blacktriangle\hspace{.4ex}}%
    {4}{\hspace{.5ex}\blacksquare\hspace{.5ex}}%
  }[\hspace{.4ex}\overset{\mbox{\tiny#1}}{\CIRCLE}\hspace{.5ex}]%
}

\newcommand{\ieq}{= \hphantom{+}} 
 
\newcommand{\ieqb}{\hphantom{+} &= \hphantom{+}}

\newcommand{\iti}{\mathrel{\hphantom{=} \times \kern-0.02em}}

\newcommand{\vxi}{\boldsymbol{\hat{\xi}}}

\newcommand{\veta}{\boldsymbol{\hat{\eta}}}

\newcommand{\vxivxi}{\vxi \otimes \vxi}
\newcommand{\vetaveta}{\veta \otimes \veta}

\newcommand{\prt}{\mathbf{p}}
\newcommand{\prtset}{\mathcal{P}}
\newcommand{\ddprt}{\dd\prt}
\newcommand{\prtprt}{\prt \otimes \prt}
\newcommand{\ddprtdeps}{\ddprt \left(u, \varphi, w\right)}

\begin{document}

\articletype{Frenkel Special Issue}

\title{Scratching a 50-year itch with elongated rods}

\author{
\name{R.~N.~P.~Creyghton\textsuperscript{a} and B.~M.~Mulder\textsuperscript{a}\thanks{CONTACT: Bela~M.~Mulder. Email: mulder@amolf.nl} }
\affil{\textsuperscript{a}Institute AMOLF, Science Park 104, 1098 XG Amsterdam, The Netherlands}
}

\maketitle

\begin{abstract}
The classical Oseen--Frank theory of liquid crystal elasticity is based on the experimentally verified fact that there are three independent modes of distortion, each with its associated elastic constant. On the other hand the arguably more first-principles order parameter-based Landau--de Gennes theory only involves two independent elastic modes. The resulting ``elastic constants problem'' has led to a considerable amount of vexation among theorists. In a series of papers at the turn of the century Fukuda and Yokoyama suggested that the resolution of this problem could be found in the proper treatment of non-local effects in the ideal part of the free energy. They used an ingenious, but technically complex, technique based on a field-theoretic approach to semi-flexible polymers. Here we revisit their idea but now in the more accessible framework of density functional theory of rigid particles. Our work recovers their main results for rod-like particles, in that generically an ordered assembly of non-interacting rods has three independent elastic constants associated to it that all scale as the square of the length of the particles and obey the inequalities $K_2 < K_1 < K_3$. We also consider the case of disk-like particles, and then find in line with expectations that $K_3 < K_1 < K_2$.
\end{abstract}


\begin{published}
This is an Accepted Manuscript of an article published by Taylor \& Francis in the Frenkel Special Issue of Molecular Physics on 18 June 2018, available online: \url{https://www.tandfonline.com/doi/10.1080/00268976.2018.1481234}.
\end{published}

\section{Introduction}
For almost 50 years, the phenomenological understanding of nematic liquid crystals (LCs) has had a proverbial elephant standing in the room. The venerable Oseen--Frank (O--F) theory \cite{Oseen_1933,Frank_1958} describes the free energy of distortion in terms of the director field $\vn(\vr)$, a vector along the local axis of uniaxial symmetry of the nematic state.  It identifies \emph{three} independent types of distortion as contributing to the free energy of distortion, and hence is characterised by \emph{three} independent elastic constants. The existence of these three independent distortion modes has been verified experimentally, and the elastic constants have been measured for many substances \cite{DeGennes-Prost, Vertogen-DeJeu}. De Gennes, however, argued that the director is an ill-defined concept, and that the proper description of the nematic phase in line with Landau's generic analysis of symmetry-breaking phase transitions \cite{Landau-Lifshitz_1980} requires the use of a second-rank tensorial order parameter $\tQ(\vr)$. This leads to the so-called Landau--de Gennes (LdG) theory \cite{DeGennes_1971}, which many would argue is the preferred continuum model of the nematic phase. Strikingly, the dictates of oriental invariance of the distortion free energy in this case only allow for \emph{two} independent distortion modes, and hence the theory only involves \emph{two} independent elastic constants. This discrepancy forms the heart of what is sometimes called the ``elastic constants problem'', and over which a significant amount of ink has been spilled, to which we are about to add. 

Of course, these phenomenological and macroscopic descriptions can be substantiated by more fundamental microscopic theories. These molecular statistical theories are based on the properties of the constituent particles and their interactions. Generically they involve two, competing terms in an expression for the free energy. One term represents the ideal entropy, favouring (positional and orientational) disorder of the mesogenic particles. The other term term accounts for the interactions, be they entropic or enthalpic in nature, and favours orientational order at higher densities or lower temperatures \cite{Onsager_1949,Maier-Saupe_comb}. It is in the latter term that many authors have sought the microscopic origin of the Frank elastic constants \cite{Nehring-Saupe_1971,Priest_1973,Straley_1973,Poniewierski-Stecki_1979}. This was successful in that approximate expressions for the constants were derived, but did not resolve the discrepancy with de Gennes' two constants \cite{Lubensky_1970,Poniewierski-Sluckin_1985}. Moreover, subtle issues arising from mapping essentially \emph{non-local} quantities to terms in the \emph{local} free energy \cite{Yokoyama_1997} remained.

Almost 20 years ago, Fukuda and Yokoyama (F\&Y) in a series of papers presented a completely novel viewpoint on the non-local to local mapping problem and its consequences for the distortion free energy by taking into account not just the position and orientation of mesogenic particles, but also the fact that they are of finite length \cite{Fukuda_1999,Fukuda-Yokoyama_2001a,Fukuda-Yokoyama_frank}. They modelled the latter aspect by starting from a polymeric description of the mesogens, and passing to the limit of infinite rigidity to obtain a reduced description in terms of a single orientation per particle. Surprisingly, their approach showed that the ideal part of the free energy by itself can give rise to three independent distortion modes if one self-consistently accounts for the way that the entropic cost of local reorientations in the nematic are propagated over small but finite distances due to the finite length of the particles. It is fair to say this work received only scant attention. This is probably partly due to the slightly ``esoteric'' nature of their problem formulation, but definitely also to the daunting technicalities involved in their field-theoretic calculations. 

Here we would like to rekindle interest in the intriguing observations of F\&Y by revisiting this problem, using a direct and  hopefully more accessible approach. We do this by assuming that the particles we are dealing with are rigid from the outset. This has a number of advantages. First and foremost, it makes direct contact with a large body of existing literature on statistical theories of LCs, which typically assume that the mesogens are rigid. Next, it both conceptually and practically simplifies the calculations involved. At the same time, it also addresses the subtle issue in statistical mechanics first raised by Van Kampen \cite{vanKampen_1981} in the context of molecular conformations, whether the limit of infinitely hardening a soft constraint yields the same thermodynamical behaviour as taking the constraint to be holonomic (= rigidly fixed) at the outset. Finally, our approach is readily extended to deal with different particle shapes, which we illustrate by treating disc-like particles.

The outline of this article is as follows. First we give a brief summary of the three relevant theoretical frameworks: Oseen--Frank, Landau--de Gennes, and density functional theory (DFT) for nematics, and in particular trace the history of the elastic constants problem. Next, we indeed scratch the 50-year itch, by sketching our calculation in some detail. We start with the formulation of a DFT that accounts for the finite length of rigid rods in the ideal term of the free energy. After an expansion in the length of the particles, we arrive at expressions for the elastic constants. We then show how are approach can also be used to obtain results for disc-like particles. Finally, we will elaborate on the nature of entropy-based elasticity and the possibilities of measuring or simulating it.

\subsection{Theories of elasticity in nematics}
The Oseen--Frank theory applies to nematics that are homogeneous both in density $\rho$ and in the degree of orientational order. What varies -- and smoothly so -- is just the local average orientation of the mesogens $\vn(\vr)$. The system’s bulk free energy is written as
$
F(\vr) = \int \ddr \, \left\{f_{0}(\vr) + f\td{d}(\vr) \right\},
$
where the first term $f_{0}(\vr)$ is the free energy density of the homogeneously aligned nematic and the second term the distortion free energy density, which is explicitly given by 
\begin{align}
f_{\text{d\tiny{\,(O-F)}}}(\vr) = \tfrac{1}{2} K_{1} (\mathrm{div}~\vn)^{2} + \tfrac{1}{2} K_{2} (\vn \cdot \mathrm{curl}~\vn)^{2} + \tfrac{1}{2} K_{3} (\vn \times \mathrm{curl}~\vn)^{2}. \label{eq:frankfd}
\end{align}
Here $K_{1},K_{2}$ and $K_{3}$ are the three elastic constants that characterise the resistance against \emph{splay}, \emph{twist} and \emph{bend} deformations respectively. These are (implicitly) dependent on density and temperature, partly through the degree of ordering. In Appendix~\ref{sec:frankrules} we recapitulate the derivation and provide a set of useful grouping and simplification rules in terms of a convenient linear combination of the constants, named $k_1,k_2,k_3$. Note that the free energy density is invariant under the inversion $\vn(\vr) \rightarrow -\vn(\vr)$, which respects the apolar nature of the nematic state. 

Since $f\td{d}$ is an energy density and $(\nabla \vn)^{2}$ has as dimension length${}^{-2}$, the $K_{i}$ must be an energy per unit length. Their order of magnitude is comparable to the interaction energy per molecular dimension scale, which is about $10^{-12}$ J/m or pico-Newtons for typical nanometre-sized organic  thermotropic mesogens \cite{DeGennes-Prost}. Typical experimental values obey $K_2 \leq K_1 \leq K_3$ \cite{Ericksen_1966}. Even though the constants scale with temperature or the order parameter, their ratios are relatively fixed for many given systems.

While phenomenologically successful, the O--F theory is based on the hard to formally define notion of the director field, and is restricted to systems with homogeneous degree of ordering, failing e.g.\ to properly deal with boundary-induced singularities in the orientation field. Moreover, by the Landau criteria \cite{Landau-Lifshitz_1980}, the order parameter should carry an irreducible representation of the rotation group, but respect the aforementioned macroscopic a-polarity. This led de Gennes to introduce the symmetric and traceless second-rank tensor order parameter $\tQ(\vr)$. This order parameter can be given a microscopic interpretation in terms of an ensemble average. To that end consider a unit vector $\vom$ along the long axis of a mesogen, and the normalised local orientational distribution function (ODF) $\psi(\vr,\vom)$. This allows us to define
\begin{equation}
\tQ(\vr) = \int \left( \vomvom - \frac{1}{3} \mathbb{1} \right) \, \psi(\vr,\vom) \, \dd \vom,
\label{eq:opOmicro}
\end{equation}
where $\otimes$ denotes the tensor product and the second constant term is in the integrand is added to ensure that the result is traceless. 
Since $\tQ(\vr)$ is symmetric, it can be diagonalised. Since it is traceless, one of its three eigenvalues can be expressed in terms of the other. If two eigenvalues are equal, $\tQ$ is called \textsl{uniaxial}, otherwise it is biaxial. In this work, we limit ourselves to the first case. For the spatially homogeneous uniaxial case we can write \cite{Gramsbergen_1986}
\begin{equation}
\tQ = S \, \vn \otimes \vn - \tfrac{1}{3} S \, \mathbb{1},
\label{eq:QMarcoUniaxial}
\end{equation}
which introduces the scalar order parameter $S$ that parametrises the degree of orientational order. It can be expressed as an ensemble average involving the angle $\theta$ between the long molecular axis and the director $\vn$, as
\begin{equation}
S = \tfrac{3}{2} \langle (\vom\cdot\vn)^2 \rangle - \tfrac{1}{2} = \langle P_{2}(\cos \theta) \rangle,
\label{eq:SandLegendre}
\end{equation}
where $P_2$ is the second Legendre polynomial. 

For assessing the impact of spatial variation in the ordering state of the LC on the free energy density, de Gennes proposed an expansion to second order in the derivatives of $\mathbf{Q}\left(\mathbf{r}\right)$, as is appropriate to capture the lowest order, long wavelength perturbations, yielding (here and later, summation over equal indices is implied):
\begin{equation}
f\td{d~(LdG)}(\vr) = \tfrac{1}{2}L_1 \, \pda Q_{\beta \gamma}(\vr) \, \pda Q_{\beta \gamma}(\vr)  + \tfrac{1}{2}L_2 \, \pda Q_{\alpha \gamma}(\vr) \,  \pdb Q_{\beta \gamma}(\vr).
\label{eq:degennesfd}
\end{equation}
The two distinct terms in this expression are, as one readily verifies, the only two independent scalar invariants one can construct that are second order both in $\tQ$ and spatial derivatives. If the order parameter is assumed uniaxial (i.e. of the form \refeq{eq:QMarcoUniaxial}) and $S$ constant, de Gennes' distortion free energy can be reduced to Frank's. If the expansion is indeed limited to the two derivative terms as in \refeq{eq:degennesfd}, then the first and third Frank constant are found to be equal. Specifically,  $K_{1} = K_{3} = 2 \left( L_{1} + \tfrac{1}{2} L_{2} \right) S^{2}$ and $K_{2} = 2 L_{1} S^{2}$ \cite{Stephen-Straley_1974}. This fact was pointed out shortly after de Gennes made his proposal \cite{Lubensky_1970}. From the late 1960s onwards, the conundrum of the ``missing'' elastic constant was discussed many times to which we will refer below.

A number of authors explored the obvious route of expanding \refeq{eq:degennesfd} to higher orders. However, de Gennes' restriction to quadratic terms only was not only physically well grounded, but also has the virtue of simplicity. Indeed, while by adding higher-order terms starting with ones of the form $Q\partial Q \partial Q$ one can lift the degeneracy of the splay and bend mode, there is a price to pay. The large number of invariant contractions of such expressions leads to a plethora of novel, independent distortion terms that are formally distinct but difficult to characterise on physical grounds \cite{Longa-ea_1987}. In spite of this many authors continue to find reason to postulate (and sometimes find some physical or mathematical use for) several additional constants in the LdG picture \cite{Poniewierski-Sluckin_1985,Longa-Trebin_1989a,Ball-Majumdar_2010,Ball-Zarnescu_2011,Mucci-Nicolodi_2012}. It is, therefore, fair to say that the ``elastic constants problem'' is alive and well.  

Meanwhile, attempts were made to not just \emph{identify} distinct elastic terms, but also to calculate their \emph{values} from first principles. This was done for several systems and with various methods, including expansions of the ODF, an extended Onsager approach and an application of DFT (i.a. Refs. \citealp{Priest_1973,Straley_1973,Poniewierski-Stecki_1979,Somoza-Tarazona_1989c}). All have in common that they explicitly consider the contribution due to the interaction between the mesogens as the source of the elastic behaviour, and that the found constants are proportional to $L^{4}$, where $L$ is the length of the particles. 

\section{DFT for finite length particles}
\label{sec:nonlocal}

We study a system of rigid cylindrically symmetric particles. The system is homogeneous in density, but can have a spatially varying degree and preferential direction of orientational order. DFT teaches us that the ideal part of the free energy is given by the expression
\begin{equation}
\beta \Fid=\rho \int \ddr \int \ddom \, \psirom \left( \log \psirom + \log \rho \thvol - 1 \right),
\label{eq:idealtermdetail}
\end{equation}
where $\rho=N/V$ is the number density and the local ODF is defined as
\begin{equation}
\psirom = \frac{\left\langle \sum_{i=1}^{N} \delta(\vr - \vr_{i}) \delta(\vom - \vomi) \right\rangle}{\left\langle \sum_{i=1}^{N} \delta(\vr - \vr_{i}) \right\rangle},
\label{eq:generalODF}
\end{equation}
which is normalised to unity when integrated over all particle orientations. 

\begin{figure}
\centering
 \vspace{-2ex}
 \includegraphics[width=.6\textwidth]{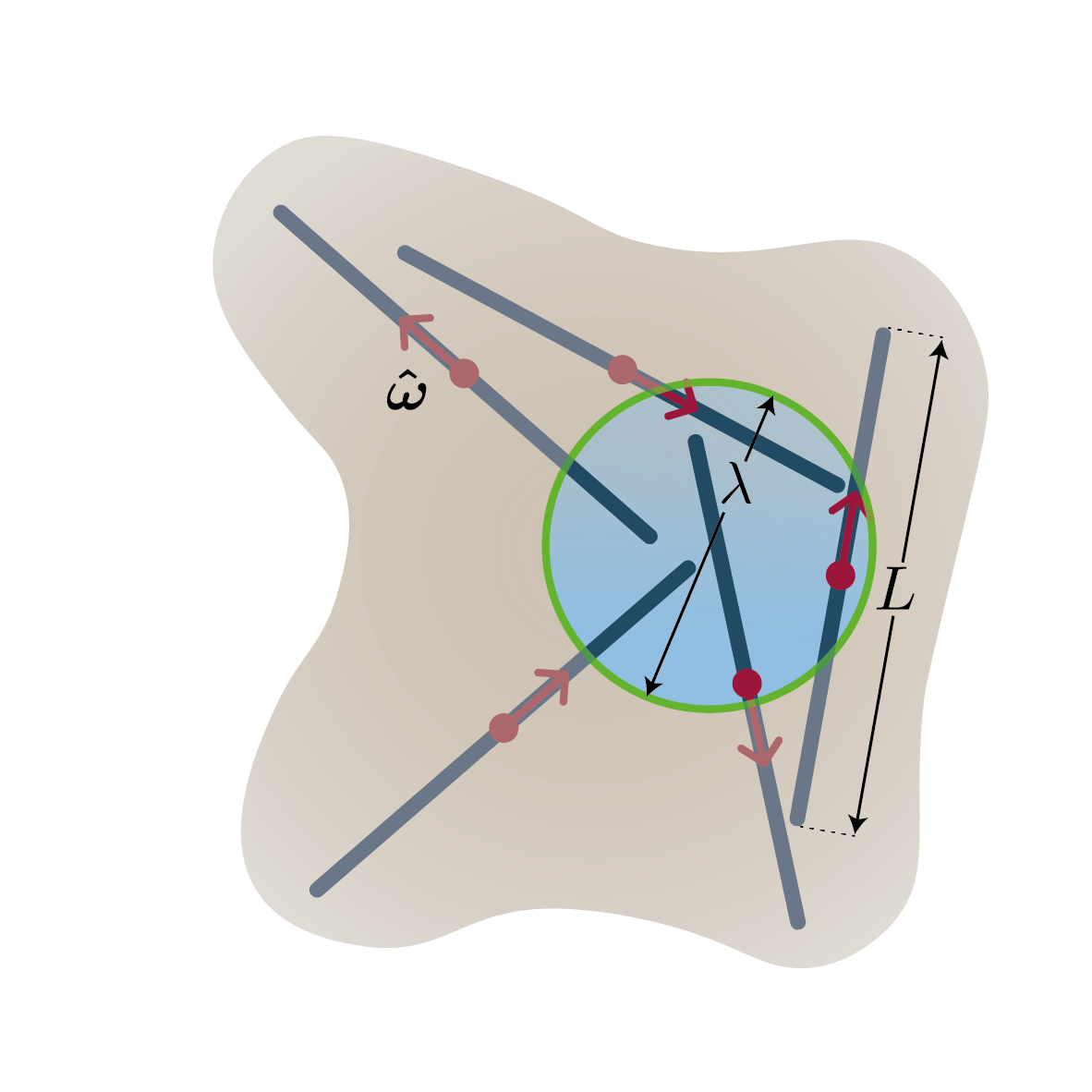}
 \vspace{-2ex}
   \caption{Pictorial explanation of non-local contributions to the ideal (entropy) term of the free energy: rods (length $L$, orientation $\vom$) with a center of mass (red dots) outside a local domain at the scale ($\lambda$) of distortions (green circle) can be relevant to it as well.}
   \label{fig:nonlocal}
\end{figure}

If particles are of finite length, changes in the centre of mass position and/or orientation of a particle will influence the local density at other locations within the particle's length scale $L$. F\&Y suggested to reformulate \refeq{eq:idealtermdetail} to account for this influence. In other words, if this novel length scale in between the system size and the localisation of centres of mass is relevant, as e.g.\ when considering arbitrary spatial variations of $\psirom$, our concept of \emph{locality} needs to be adjusted. This idea is illustrated in Fig.~\ref{fig:nonlocal}. 
This goes beyond the majority, if not all, of past and current molecular statistical approaches to simple LCs, which routinely assume that the ideal free energy density is a purely local quantity even for spatially \emph{in}homogeneous systems, an interpretation which in the proposed more fine-grained view on locality is strictly speaking only valid for point particles with internal degrees of freedom given by the ``spin'' $\vom$.

We, therefore, proceed by defining the density appropriate to the novel length scale through
\begin{align}
o(\vr,\vom)
&\equiv \left\langle \frac{1}{L}  \int_{-\frac{1}{2}L}^{+\frac{1}{2}L} \dd l \sum_{i=1}^{N} \delta(\vr - (\vr_{i} + l \vom_{i})) \delta(\vom - \vom_{i}) \right\rangle \nonumber \\
&= \rho \frac{1}{L} \int_{-\frac{1}{2}L}^{+\frac{1}{2}L} \dd l \,  \psi(\vr-l \vom,\vom) ,\label{eq:spdfwithlength}
\end{align}
to which we will refer as the $L$-density. 
Clearly, the $L$-density $o(\vr,\vom)$ contains the length-weighted contribution of all particles with orientation $\vom$ at $\vr$, irrespective of whether their centre of mass lies inside or outside an averaging volume of dimension $\lambda \ll L$. One also checks that as expected
\begin{equation}
\int \ddr \int \ddom \,  o(\vr,\vom) = \int \ddr \, \rho = N,
\label{eq:onormal}
\end{equation}
where the spatial integration runs over the full spatial volume, assumed large enough that boundary effects can be neglected. For future reference, we introduce the following shorthand notation for the average of a function which depends on the coordinate along along the length of a particle
\begin{equation}
\left[ f(\mathbf{v},l) \right]_{l} \equiv  \frac{1}{L} \int_{-\sfrac{1}{2}L}^{+\sfrac{1}{2}L} \dd l \, f(\mathbf{v},l),
\label{eq:LengthAvNot}
\end{equation}
which will ease the notation throughout. Symmetry entails $\left[ f(\mathbf{v},l) \right]_{l} = \left[ f(\mathbf{v},-l) \right]_{l} $. Using the newly defined $L$-density, we define the second rank $L$-order tensor
\begin{align}
\tO(\vr) &= \int \ddom \, o(\vr,\vom) \, \vomvom .
\label{eq:deftensordO}
\end{align}
Note that we forgo to make this tensor traceless at this point, which simplifies the algebraic computations further on. The constant multiple of the identity tensor needed to produce the true order parameter can always be added on when required. 

We now wish to evaluate the ideal free energy \refeq{eq:idealtermdetail} under the constraint of an imposed $L$-order tensor field $\tO(\vr)$, in order to determine the ideal contribution to the distortion free energy. To that end we introduce a tensorial Lagrange parameter field $\tH(\vr)$ and perform the minimisation
\begin{align}
\frac{\delta}{\delta \psirom} \left\{ \beta \Fid [\psi] - \beta \mu \rho \int \ddr \int \ddom \, \psirom + \beta \int \ddr \, \tH(\vr) \, \colon \tO(\vr) \right\} = 0,
\label{eq:freeconstrained}
\end{align}
where the full colon $:$ denotes double contraction over all indices and the chemical potential $\mu$ is used as usual to enforce the normalisation of $\psirom$. We now note that
\begin{align}
\int \ddr \, \tH(\vr) \, \colon \tO(\vr)  &=\rho \int \ddr \int \ddom [\psi(\vr-l\vom,\vom)]_l \tH(\vr):\vomvom \nonumber \\
&= \rho \int \ddr \int \ddom \, \psi(\vr,\vom) [\tH(\vr-l\vom)]_l:\vomvom,
\end{align}
where we have used the linearity and invariance with respect to the direction of integration of the averaging operator $[.]_l$, and neglected boundary terms generated by the shift of the origin of spatial integration. This latter form readily allows the functional derivative of \refeq{eq:freeconstrained} to be taken. The result is
\begin{equation*}
\log \psirom + \log \thvol \rho - \beta \mu + \beta  \left[ \tH(\vr - l \vom) \right]_{l} \vomvom =0,
\end{equation*}
leading, after elimination of the chemical potential, to
\begin{align}
\psirom &= \frac{N}{\Z[\tH] } e^{ -\beta [\tH(\vr - l \vom)]_{l} \, \colon \vomvom}, \label{eq:psiromnormal}
\intertext{where the normalization factor is given by}
\Z[\tH] &= \rho \int \ddr \int \ddom \, e^{-\beta [\tH(\vr - l \vom)]_{l} \, \colon \vomvom}. 
\label{eq:partitionfunc}
\end{align}
Substituting the explicit form \refeq{eq:psiromnormal} into \refeq{eq:idealtermdetail}, we obtain the ideal free energy now as a functional of the conjugate field
\begin{equation}
\beta \Fid[\tH] = -\beta \int \ddr \,  \tH(\vr) \, \colon \tO (\vr) - N \log \Z[\tH]  + N\left(\log \thvol \rho + \log N - 1 \right),
\label{eq:Fid3}
\end{equation}
Minimising this form with respect to the field, we recover the self-consistency condition that fixes $\tH(\vr)$:
\begin{align}
 \tO(\vr) &=  - \frac{N}{\beta} \frac{\delta \log \Z[\tH]}{\delta \tH(\vr)} \nonumber \\
 &= N \frac{\int \ddom \, \left[ e^{-\beta [\tH(\vr + (l\pr \hspace{-0.25ex}- l) \vom)]_{l} \ctf{2} \vomvom} \right]_{l\pr} \, \vomvom}{\int \ddr\pr \int \ddom \, e^{-\beta [\tH(\vr\pr - l \vom)]_{l} \, \colon \vomvom}},
 \label{eq:Ostartdirect}
\end{align}
which one would also obtain by inserting \refeq{eq:psiromnormal} directly into the definition \refeq{eq:deftensordO} with the help of \refeq{eq:spdfwithlength}.
Our goal of obtaining the distortion free energy $\beta \Fid[\tO]$ in terms of the imposed ordering field is reached if we would be able to invert \refeq{eq:Ostartdirect} and eliminate $\tH(\vr)$ in favour of $\tO(\vr)$. In general this appears intractable due to the non-linearities involved. However, taking our cue from the analogous approach taken by F\&Y, we will show in the following that we can perform the necessary inversion perturbatively in the particle length $L$.

\section{Perturbative solution}
%
%
\subsection{General structure}
\label{sec:perturbativegeneral}
To obtain a perturbative solution to the self-consistency equation \refeq{eq:Ostartdirect} aimed at eliminating the field $\tH(\vr)$, we expand this field to second order on the length $L$ of the particles
\begin{equation}
\tH(\vr) = \tH\nnz(\vr) + L^{2} \tH\nnt(\vr) + \boh(L^{4}).
\label{eq:Hexpans}
\end{equation}
Note that because we assume the particles to be cylindrically symmetric and inversion symmetric, odd powers of $L$ do not occur. Next, this expansion is introduced in the right-hand side of \refeq{eq:Ostartdirect}, which is then expanded in turn to allow an order-by-order solution. Here we only broadly sketch the salient elements of the necessary derivation, referring the interested reader to the relevant appendices for the details of the, at times algebraically involved, calculations. For ease of notation, we have also tacitly set $\beta=1$ throughout.

The first step involves the expansion of the exponents in the integrands on the right-hand side of \refeq{eq:Ostartdirect}. We find for the numerator and denominator, respectively:
\begin{align}
\left[ e^{-[\tH(\vr + (l\pr \hspace{-0.25ex}- l) \vom)]_{l} \ctf{2} \vomvom} \right]_{l\pr} &= e^{-\tH\nnz(\vr) \ctf{2} \vomvom }\left( 1+L^2 N \nnt (\vr,\vom) \right) \label{eq:numexpand}\\
e^{- [\tH(\vr - l \vom)]_{l} \ctf{2} \vomvom} &= e^{-\tH\nnz(\vr) \ctf{2} \vomvom }\left( 1+L^2 D\nnt(\vr,\vom) \right), \label{eq:denexpand}
\end{align}
where
\begin{align}
N \nnt (\vr,\vom) = -\tH\nnt(\vr)\ctf{2}\vomvom+\frac{1}{24}\Big(& \left(\vom \cdot \nabla \otimes \tH\nnz(\vr)\ctf{2}\vomvom \right)^2 \label{eq:defN}\\
&-2 \, \vomvom\ctf{2}\nabla\otimes\nabla\otimes\tH\nnz(\vr)\ctf{2}\vomvom \Big) \nonumber \\
D\nnt(\vr,\vom)  =  -\tH\nnt(\vr)\ctf{2}\vomvom -\frac{1}{24}\Big(& \vomvom\ctf{2}\nabla\otimes\nabla\otimes\tH\nnz(\vr)\ctf{2}\vomvom\Big). \label{eq:defD}
\end{align}
We now define a useful set of integrals:
\begin{align}
 M_{\nu_{1} \dots \nu_{q}}(\vr) &=  \int \ddom \, e^{- \tH^{(0)}(\vr) \, \colon \vomvom } \omega_{\nu_{1}} \dots \omega_{\nu_{q}} \label{eq:defMqelems} \\
M_{\nu_{1} \dots \nu_{q}}^{[1,1]}(\vr) &= \int \ddom \, e^{- \tH^{(0)}(\vr) \, \colon \vomvom } \left(\omcia \pda H_{\sigma\tau}^{(0)}(\vr)\omcis\omcit \right)\left(\omcib\pdb H^{(0)}_{\gamma \delta}(\vr) \omcig\omcid \right) \omega_{\nu_{1}} \dots \omega_{\nu_{q}} \label{eq:defMqdHdHelems}  \\
M_{\nu_{1} \dots \nu_{q}}^{[2]}(\vr) &= \int \ddom \, e^{- \tH^{(0)}(\vr) \, \colon \vomvom}\left( \omcia\omcib\partial^{2}_{\alpha\beta} H_{\sigma\tau}^{(0)}(\vr)\omcis\omcit\right)\omega_{\nu_{1}} \dots \omega_{\nu_{q}} \label{eq:defMqddHelems}
\end{align}
In terms of these integrals and combining \refeq{eq:Ostartdirect}, with the expansions \ref{eq:numexpand}\&\ref{eq:denexpand} and the definitions \ref{eq:defN}\&\ref{eq:defD} and neglecting higher orders in $L^2$, we can write
\begin{equation}
O_{\sigma \tau}(\vr) =N \frac{M_{\sigma \tau}(\vr)+L^{2} \left( -H_{\alpha \beta}\nnt(\vr) M_{\alpha \beta \sigma \tau}(\vr) + \frac{1}{24} \left\{M_{\sigma \tau}^{[1,1]}(\vr)- 2 M_{\sigma \tau}^{[2]}(\vr) \right\}\right)}{\int \ddr\pr \Big\{M(\vr\pr) - L^2 \left( H\nnt_{\alpha\beta}(\vr\pr) M_{\alpha\beta}(\vr\pr) + \frac{1}{24} M_{}^{[2]}(\vr\pr)\right)\Big\}}.
\label{eq:Ofraction}
\end{equation}
As the imposed orientational ordering field on the left-hand side of \refeq{eq:Ofraction} is by definition independent of $L$, we can immediately read off the zeroth-order equation:
\begin{equation}
O_{\sigma \tau}(\vr) =N \frac{M_{\sigma \tau}(\vr)}{\int \ddr\pr M(\vr\pr)}.
\label{eq:order0}
\end{equation}
Due to our assumption that the density of the system is spatially homogeneous, we have that
\begin{equation}
\Tr \tO(\vr) = \int \ddom \, o(\vr,\vom) = \rho.
\end{equation}
This implies that $M(\vr) = \Tr M_{\sigma\tau}(\vr)$ is in fact a constant, which we call $M_{0}$. Solving \refeq{eq:order0} will determine the zeroth-order contribution to the effective field $\tH\nnz(\vr)$. 

Since the denominator of \refeq{eq:Ofraction} is a scalar, it is sufficient to require
\begin{equation}
-H_{\alpha \beta}\nnt(\vr) M_{\alpha \beta \sigma \tau}(\vr) + \frac{1}{24} \left\{M_{\sigma \tau}^{[1,1]}(\vr)- 2 M_{\sigma \tau}^{[2]}(\vr) \right\} =0 \label{eq:order2}
\end{equation}
to ensure that $\tO(\vr)$ has no tensorial components of order $L^2$. This equation, which implicitly depends on the previously determined $\tH\nnz(\vr)$, can be used to obtain the second-order contribution to the effective field $\tH\nnt(\vr)$.

Since the normalisation of $\tO(\vr)$ precludes terms of order $L^2$ in a scalar pre-factor, of necessity the second-order term in the denominator of \refeq{eq:Ofraction} must also vanish. That this is indeed the case follows from the observation that
\begin{equation}
\nabla\cdot\nabla M(\vr) = M^{[1,1]}(\vr)-M^{[2]}(\vr).
\label{eq:MtoSurfaceIdentity}
\end{equation}
Being a divergence, the integral over this quantity is actually a surface term, which we can neglect in the infinite volume limit. Any multiple of this term can, therefore, be freely added to the integrand in the denominator, yielding the identity
\begin{align}
& \int \ddr\pr\,\left( H\nnt_{\alpha\beta}(\vr\pr) M_{\alpha\beta}(\vr\pr) + \frac{1}{24} M_{}^{[2]}(\vr\pr) \right) \nonumber \\
= &\int \ddr\pr \,\left( H\nnt_{\alpha\beta}(\vr\pr) M_{\alpha\beta}(\vr\pr) + \frac{1}{24}\left\{  2M_{}^{[2]}(\vr') -M^{[1,1]}(\vr\pr)\right\} \right) \nonumber \\
= &-\int \ddr  \Tr\left( -H_{\alpha \beta}\nnt(\vr) M_{\alpha \beta \sigma \tau}(\vr) + \frac{1}{24} \left\{M_{\sigma \tau}^{[1,1]}(\vr)- 2 M_{\sigma \tau}^{[2]}(\vr) \right\} \right) =0 .\label{eq:order2iden}
\end{align}
Below we will explicitly determine these solutions under the assumption that the imposed orientational ordering is uniaxially symmetric and of equal degree everywhere, a case we propose to call the Frank state.

%
%
\subsection{The Frank state}
\label{sec:frankstate}
In the the Frank state the imposed $L$-local ordering is characterised by an order parameter of the type we already encountered in \refeq{eq:QMarcoUniaxial}. Using the definition \refeq{eq:deftensordO} we see that this implies
\begin{equation}
\tO(\vr)= \rho \left\{ S\, \vn(\vr) \otimes \vn(\vr)-\frac{1}{3}(S-1) \, \mathbb{1} \right\} .\label{eq:OinFrank}
\end{equation}
The zeroth-order equation to be solved thus is
\begin{equation}
\rho \left\{S \, n_{\sigma}(\vr)n_{\tau}(\vr)-\frac{1}{3}(S-1) \,\delta_{\sigma\tau} \right\} = \rho \frac{M_{\sigma\tau}(\vr)}{M_{0}}.
\label{eq:self0}
\end{equation}
We now note that as the left-hand side is expressed on the the tensor basis $\vn(\vr)\otimes\vn(\vr)$ and $\mathbb{1}$, $\tH^{(0)}(\vr)$ must also be expressible on this basis
\begin{equation}
\tH^{(0)}(\vr)=U \, \vn(\vr) \otimes \vn(\vr)-\frac{1}{3} U \,\mathbb{1}, \label{eq:HinFrank}
\end{equation}
i.e.\ sharing the same local the uniaxial symmetry axis as $\tO(\vr)$. Choosing $\tH\nnz(\vr)$ to be traceless is a matter of choice, as any adding a constant multiple of the identity does not influence the end results. Using this parametrisation we can show (cf. Appendix \ref{sec:listingdnterms} and \ref{sec:orientationalintegrals}) that
\begin{align}
M_{0} &= J_{0,0} \nonumber \\
M_{\sigma\tau}(\vr) &= J_{0,2} n_{\sigma}(\vr)n_{\tau}(\vr)+J_{2,0}\delta_{\sigma\tau}, \label{eq:MasJdn}
\end{align}
where
\begin{align}
J_{0,0} &= \frac{2 \pi ^{3/2} e^{U/3} \,\text{erf}\left(\sqrt{U}\right)}{\sqrt{U}} \label{eq:partizero} \\
J_{2,0} &= \frac{e^{-2 U/3} \left(e^U \pi ^{3/2} (2 U-1) \,\text{erf}\left(\sqrt{U}\right)+2 \pi 
   \sqrt{U}\right)}{2 U^{3/2}}  \\
J_{0,2} &= - \frac{e^{-2 U/3} \left(e^U \pi ^{3/2} (2 U-3) \,\text{erf}\left(\sqrt{U}\right) + 6 \pi 
   \sqrt{U}\right)}{2 U^{3/2}}.
\end{align}
Inserting these expressions into \refeq{eq:self0}, we are left with the non-linear self-consistency condition
\begin{equation}
S = \frac{1}{4} \left(-\frac{6 \, e^{-U}}{\sqrt{\pi U} \, \text{erf}\left(\sqrt{U}\right)}+\frac{3}{U}-2\right).
\label{eq:SfromUzero}
\end{equation}
Although analytically inverting this relation appears impossible, the numerical solution can readily be found, and we will in the following simply denote it by $U$. One verifies the properties $U=0$ when $S=0$, $U\rightarrow \infty$ as $S\rightarrow -\sfrac{1}{2}$, and $U\rightarrow -\infty$ as $S\rightarrow 1$. Note that $\text{erf}(\sqrt{U})/\sqrt{U}$ remains real even for negative values of $U$. 

We then turn to the second-order equation \eqref{eq:order2}. We again note all terms are second rank tensors, which by symmetry necessarily must also be expressible on the basis $\vn(\vr)\otimes\vn(\vr)$ and $\mathbb{1}$, as both $\tO(\vr)$ and $\tH^{(0)}(\vr)$ are expressed in this basis. It, therefore, suffices to take $\tH^{(2)}(\vr) = V \, \vn(\vr)\otimes\vn(\vr) + W \, \mathbb{1}$, where the two unknowns $V$ and $W$ are used to fix the two degrees of freedom. The explicit expressions for $V$ and $W$ are found in Appendix~\ref{sec:fullsolutions}.

We are now in a position to evaluate the impact of the imposed ordering profile on the ideal free energy to second order in the length of the particles. As a first step we consider
\begin{equation}
\mathcal{Z}[\tH]= NJ_{0,0}+ L^2 \int \ddr \left( H\nnt_{\alpha\beta}(\vr) M_{\alpha\beta}(\vr) + \frac{1}{24} M^{[2]}(\vr) \right),
\end{equation}
where we recognise the integrand in parentheses in the second-order term as the one which, as we have shown above, vanishes identically due to the self-consistency condition. Referring back to \refeq{eq:Fid3} we can then write
\begin{equation}
\beta \Fid[\tH] = N\left(-\frac{2}{3} U S - \log J_{0,0}\right) -L^2 \int \ddr \,  \tH^{(2)}(\vr) \, \colon \tO (\vr) + N\left(\log \thvol \rho + \log N - 1 \right). \label{eq:Fidexpand}
\end{equation}
We immediately notice that apart from the default entropic cost of raising the degree of orientational order to the specified value of $S$, which is embodied in the first two terms, there is now an explicit second-order term accounting for the cost of spatially distorting the homogeneous aligned state.

%
%
\subsection{The ``ideal'' elastic constants}
\label{sec:idealconstants}
\begin{figure}
\centering
\vspace{-4ex}
 \includegraphics[width=0.75\textwidth]{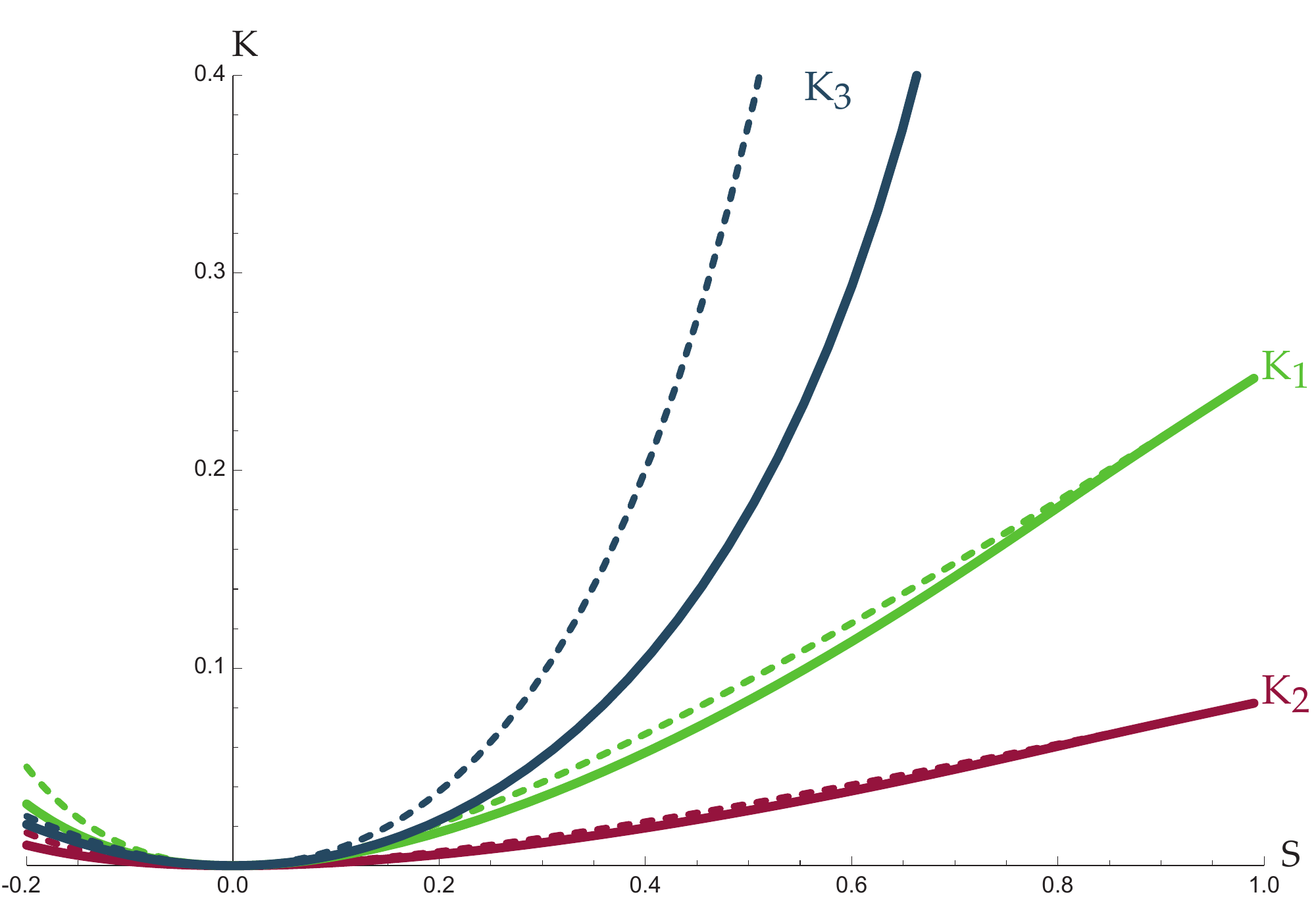}
 \vspace{-2ex}
   \caption{Frank elastic constants, found from the orientational entropy of rigid rods with finite length $L$, as a function of the scalar order parameter S. Plotted are the results from both F\&Y \cite{Fukuda-Yokoyama_frank} (dashed lines) and this work (solid lines). Units $\rho = L = \beta = 1$. The zeroth-order solution \eqref{eq:SfromUzero} was inverted numerically to obtain this plot.}
   \label{fig:constantsplot}.
  \vspace{-5ex}
\end{figure}

We can now extract the elastic constants associated with the distortion free energy implied by the $L^2$ term in \refeq{eq:Fidexpand}, by using the results on $\tH\nnt(\vr)$. To ease the notation, we drop the explicit dependence on $\vr$ of the director field. We first find using the explicit results for $V$ and $W$ (see Appendix~\ref{sec:fullsolutions}):
\begin{align}
- L^{2} \, \tH\nnt(\vr) \ctf{2} \tO(\vr)\hspace{0.5ex}
=\hspace{1ex}\frac{1}{144} &\rho L^{2} \big(2U + S(15-2U)\big) \partial_{\alpha} n_{\alpha}  \partial_{\beta} n_{\beta} \nonumber \\
+\frac{1}{288} &\rho L^{2} \big(2U + S(15-2U)\big)  (\partial_{\alpha} n_{\beta})^2 \label{eq:sol2} \\
-\frac{1}{288} &\rho L^{2} \big(10U + S(75+14U)\big) n_{\alpha}  n_{\beta}  \partial_{\alpha} n_{\gamma}  \partial_{\beta} n_{\gamma} . \nonumber
\end{align}
Using result \eqref{eq:frankK1}--\eqref{eq:frankK3} of Appendix~\ref{sec:frankrules}, restoring dimensions with $\beta= 1 / k_{\mathrm{B}} T$, and --again-- using $U=U(S)$ as implicitly given by \refeq{eq:SfromUzero}, we can then read off the elastic constants:
\begin{align}
K_1 \:&= \:\frac{1}{\;48\:} \frac{\rho}{\beta} L^2 \big(2 U + S (15-2U)\big) \label{eq:finalK1} \\
K_2 \:&= \:\frac{1}{144} \frac{\rho}{\beta}  L^2 \big(2 U + S (15-2U)\big) \label{eq:finalK2} \\
K_3 \:&=-\frac{1}{36} \frac{\rho}{\beta}  L^2 \big(2 U + S (15+4U)\big)\label{eq:finalK3}.
\end{align}
We plot these as a function of $S$ in Figure \ref{fig:constantsplot}, comparing them to the results presented by F\&Y. 
The following observations are in order. First of all, we have independently verified that the hypothesis pioneered by F\&Y that for arbitrary spatial variations of an imposed uniaxial orientational ordering field, there are three independent contributions to free energy of distortion already at the level of the ideal part of the free energy, and that these scale as the square of the length of the particle. We also recover the following results obtained by F\&Y: (i) the three ``ideal'' elastic constants obey the inequalities $K_2 < K_1 < K_3$, (ii) two of the constants $K_1 = 3 K_2$ have a fixed ratio, and (iii) $K_3$ diverges for $S\rightarrow 1$, albeit less strongly than the corresponding F\&Y result. Finally, although for $K_1$ and $K_2$ the two results are close to those of F\&Y, and even coincide in the limit $S \rightarrow 1$, the two sets of results are clearly not identical. This difference is already apparent in the limit of weak ordering ($S \ll 1$), essentially the limit in which the LdG approach should be correct. Here we find $K_1=K_3=\frac{15}{28} S^2$, the latter two identical as expected, and $K_2=\frac{5}{28} S^2$. This should be compared with the values $K_1=K_3=\frac{3}{4} S^2$ and $K_2 =\frac{1}{4} S^2$ reported by F\&Y. 

%
%
\subsection{Generalisation to discs}
\begin{figure}
\centering
\vspace{-4ex}
 \includegraphics[width=0.75\textwidth]{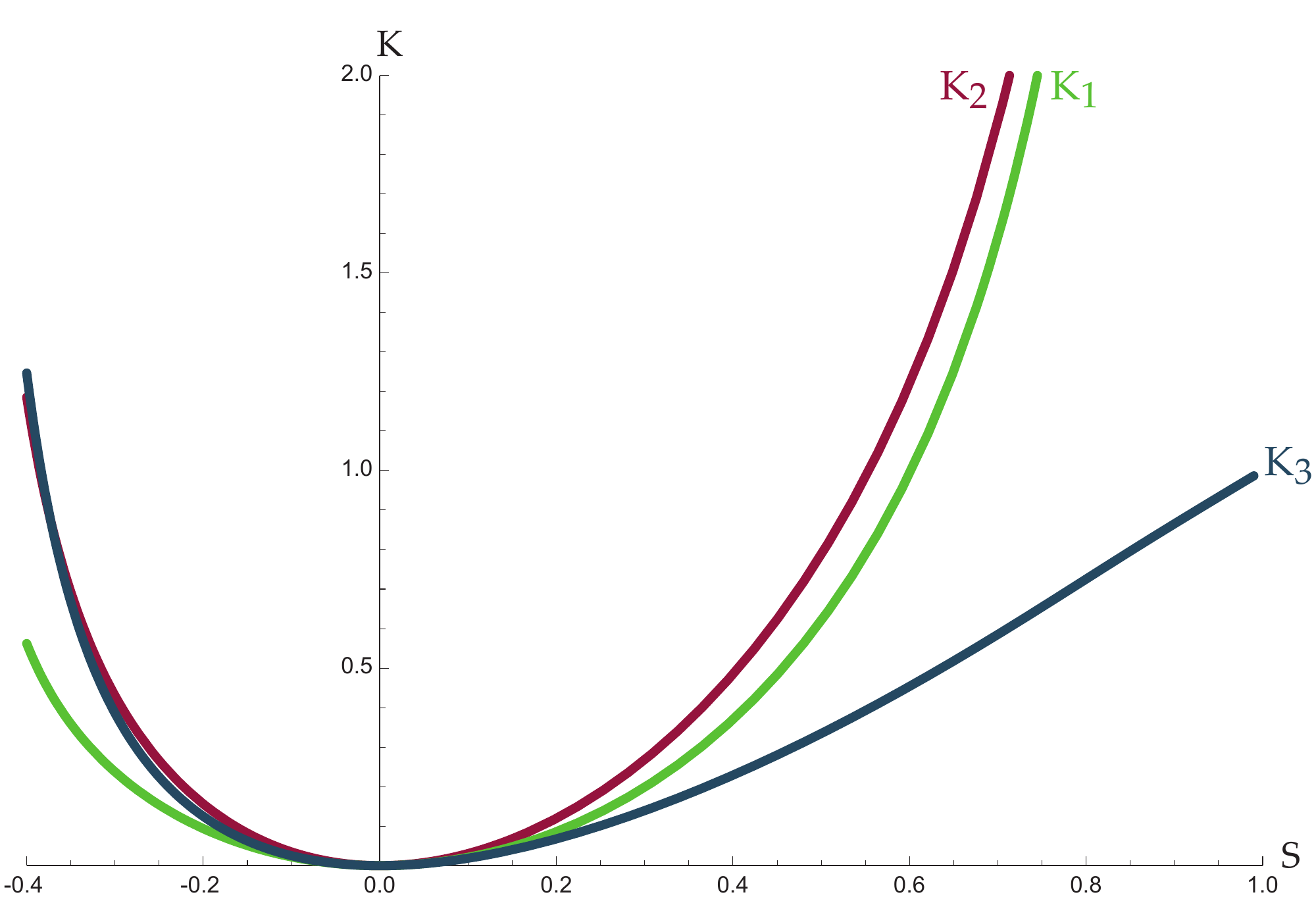}
 \vspace{-2ex}
   \caption{Frank elastic constants from the orientational entropy of rigid discs with radius $R$, as a function of the scalar order parameter S. Like in Fig.~\ref{fig:constantsplot}, units $\rho = L = \beta = 1$, and \eqref{eq:SfromUzero} was inverted numerically. Note the change in vertical scale with respect to the previous plot, and the inversion of the order of the constants.}
   \label{fig:constantsplotdiscs}.
  \vspace{-5ex}
\end{figure}

The above derivation and results for rigid rods are in fact readily generalised to any type of particle $\prtset$ whose shape possesses both axial and inversion symmetry. In Appendix~\ref{sec:discs} we show in general that we can replace the $L$-density \eqref{eq:spdfwithlength} with the $\prtset$-density
\begin{align*}
o_{\prtset}(\vr,\vom) \equiv \rho \big[  \psi(\vr - \prt,\vom) \big]_{\prtset}
\end{align*}
where the integration now runs over the volume of the particle. This properly accounts for the volume-weighted contributions of all relevant particles to the local density. In case of cylinder-shaped particles this leads to an additional term in the free energy of the form  $R^2  \tH\nn{2,R}(\vr) \ctf{2} \tO(\vr)$, where $R$ is the radius of the cylinder, next to the one of order $L^2$ already discussed before. This contribution can be obtained by collecting terms of second order in $R$ in the self-consistency conditions --- namely \refeq{eq:order2Req}, a variant of \refeq{eq:order2}. Characteristic differences with the contribution due to the length of the particles are an additional factor of $3$, essentially a reflection of the fact that the moment of inertia of a disk is $3$ times that of a rod in terms of the defining magnitude ($R$ vs. $L$), and the presence of the ``transverse'' tensor $(\mathbb{1} - \vomvom)$ within the $M$-integrals, rather than the ``axial'' $\vomvom$. 
Going through the motions of Secs.~\ref{sec:frankstate} \& \ref{sec:idealconstants}, that is: solving this equation in the Frank limit and collecting the three types of elastic terms, we arrive at the following entropic contributions to the elastic constants from a non-zero particle radius $R$:
\begin{align}
K_1\nn{R} \:&= -\frac{1}{16} \frac{\rho}{\beta} R^2 \big(2 U + S (15 + 6U)\big) \label{eq:finalK1R} \\
K_2\nn{R} \:&= -\frac{1}{48} \frac{\rho}{\beta}  R^2 \big(2 U + S (15 + 22U)\big) \label{eq:finalK2R} \\
K_3\nn{R} \:&= \;\frac{1}{12} \frac{\rho}{\beta}  R^2 \big(2 U + S (15 - 2U)\big) \label{eq:finalK3R}.
\end{align}
These are plotted as a function of scalar order $S$ in Figure~\ref{fig:constantsplotdiscs}. 
We note the following two salient features: (i) the order of the constants in magnitude is exactly reversed with respect to the rod case, which is to be expected from the subtraction in \refeq{eq:order2Req}, and (ii) both $K_1$ and $K_2$ now diverge as $S\rightarrow 1$, while $K_3$ remains finite, also exactly opposite to the rod case. Strikingly, this phenomenology precisely matches the one observed in very recent simulation results for the elastic constants for short aspect ratio ($0.2$) oblate hard ellipsoids, compared to prolate ones of aspect ratio $5$ \cite{Heymans_2017}. Finally, we again check that the first and third elastic constant are indistinguishable in the weak ordering limit $S \ll 1$: to lowest order $K\nn{R}_1=K\nn{R}_3=\frac{15}{7} S^2$, whereas $K\nn{R}_{2}=\frac{45}{15} S^2$.

\section{Discussion}
There are a number of notable aspects to the results we have presented here. First, and foremost, we independently and fully corroborate the Ansatz of F\&Y that by properly taking into account the entropic contribution of deforming an aligned nematic LC on the length scale of the particles by itself leads to the identification three independent elastic constants. Next, our standard DFT-based approach shows that the ``detour'' through the theory of linear polymers that F\&Y took towards their results is in no way essential. In fact, we can in our framework readily consider particles of any shape, and not just rods. Perhaps even more surprisingly, the results on the relative ordering in magnitude between the elastic constants appears both for rods and discs to recapitulate the ones observed in simulations with interacting particles. Together with the result that these ``ideal'' elastic constants scale as the square of the relevant dimension, be it $L^2$ for rods or $R^2$ for discs, rather than the fourth powers, which follow from considering the contributions due to the interactions between the particles, this begs the question of whether these contributions can be disentangled. 

This points to the perhaps thorny question what, if anything, these elastic constants here derived mean physically, and whether they are somehow measurable e.g.\ in a simulation. In order to do so would require considering a system of ideal rod-like particles with a prescribed, spatially homogeneous degree of order. This in principle is easy to achieve by applying a homogeneous external field that couples in a non-polar fashion to the orientation of the particles. Next, we need to meet the requirement that the spatial (length) density of the particles also be uniform at length scales smaller than that of the length of the particles. This latter requirement is of course harder to achieve in a simulation with a finite number of particles. Provided these requirements are appropriately met, one could then in principle employ spectral analysis of orientational fluctuations to extract the elastic constants (see e.g.~\citeref{Allen-ea_1996}).
  
We do hope to have shown that the at first sight possibly counterintuitive idea launched by F\&Y that there is an a purely ideal contribution to the distortion free energy for non-spherical particles is readily embedded within the standard framework of DFT of rigid particles. The fact that in the case of rods our results are similar, yet not identical, to the ones of F\&Y,  indicates that indeed the procedure followed by the latter authors includes entropic contributions due to the flexibility of particles that persist in the limit of infinite rigidity. Why these differences play out the way that they do -- significantly different $K_3$, yet almost equal $K_1$ and $K_2$ -- is unfortunately difficult to address at this point. One can, of course, argue that in reality molecules are never fully rigid, so that perhaps the F\&Y approach is more realistic. However, we should point out that their approach would be very hard to extend to other particle shapes than rods. 

Finally, here we limited our explicit calculations to the Frank state. We remark, however, that the approach as described can be generalised to the case were both the local length density and the local degree of orientational order are variable. This opens the door for studying the interesting physics of highly confined LCs, where the typical dimensions of the confining volume are comparable to the length of the particles. Recent work has shown that in such systems the competition between bulk ordering and boundary effects can give rise to novel defect-mediated orientational patterns \cite{Lewis-ea_2014,Garlea-ea_2016}, which are beyond the ken of the either the Frank--Oseen or the Landau--de Gennes approaches. Since the ratio of the length of the particles to the system dimensions is manifestly salient in these cases, one could hope that a perturbative approach as described here could be used to study these novel phenomena.

\section*{Acknowledgements}
BM would like to thank Daan Frenkel, co-supervisor of his thesis, friend and later colleague at AMOLF, who has always inspired him to think deeply about liquid crystals: this work builds on the foundation laid by our collaboration over the years. Thanks is also due to Peter Pallfy-Muhoray who passed on the `elastic constants problem' bug during a visit to the Kent State Liquid Crystal Institute at the end of the last century, which has turned out to be a highly resistant strain indeed.

\newpage
\bibliographystyle{tfo}

\appendix

\section{Frank terms and rules}
\label{sec:frankrules}
Requirements for the distortion free energy density $f\td{d}$ in the Oseen--Frank theory are that (i) it is composed of derivatives of $\vn$, (ii) it is even in $\vn$ (as the the mesogens and thus the nematic orientation are supposed to be head-tail symmetric), and (iii) terms that are relevant on the boundary alone may be neglected in most cases. As a consequence of the first two rules all terms in $f\td{d}$ are of order $(\nabla \vn)^{2}$. In principle, a list of all those terms, knowledge of the divergence theorem, curl identities and the fact that $\vn$ is a unit vector are sufficient to find the famous three Frank elastic constants. This exercise is done in de Gennes \& Prost's textbook \cite{DeGennes-Prost}.

A more transparently written (but equivalent) argument was offered by i.a. Vertogen \& De Jeu \cite{Vertogen-DeJeu}. It offers rules to identify and orders types of elastic terms that are used throughout this work. They write down the first- and second-order derivatives of the director:
\begin{align*}
f\td{d~(O-F)}(\vr) &= L_{\alpha \beta}(\vr) \pda \ncib (\vr) + L_{\alpha \beta \gamma \delta} (\vr) \left( \pda \ncib(\vr) \right) \left(\pdb \ncid (\vr)  \right) \\
&\;\;+ L_{\alpha \beta \gamma} (\vr) \pda \pdb \ncig(\vr).
\end{align*}
The $L$-tensors that are used to contract the terms contain a coefficient that will contribute to the elastic constants and a tensorial part that must be composed of elements of $\vn$ and the unit tensor.\footnote{For chiral systems, the Levi-Civita tensor is necessary as well. This introduces an extra type of elastic terms, which we will not consider here.} This is an exercise in combinatorics: all allowed ways to assign indices to a product of unit vector elements and delta's need to be listed. This similar exercise returns in Appendix~\ref{sec:listingdnterms}, and we will elaborate on it there. Once the possible $L$-tensors are listed, they are contracted with their respective vector-derivative terms from the above expression. The simplification rules that we mentioned rather qualitatively above can now be enforced using a few simple rules, which we present in index notation:\footnote{Here, as in the following, we will often omit the explicit $\vr$ dependence to avoid clutter.}
\begin{flalign}
&\text{Unit vector: }\ncia \ncia =1 & \label{eq:dnruleUnit} \\
&\text{Consequence: }\ncib \pda \ncib = \tfrac{1}{2} \pda (\ncib \ncib) = 0 & \label{eq:dnruleUnitD} \\
&\text{Surface terms (divergence theorem): neglect } \pda (\ncia \pdb \ncib) \text{ and } \pda (\ncib \pdb \ncia) &\label{eq:dnruleSurface} \\
&\text{Equal up to surface term: } \ncia (\pda \pdb \ncib) = \cancel{\pda (\ncia \pdb \ncib)} - \pda \ncia \pdb \ncib. &\label{eq:dnruleEqUTsurf} 
\end{flalign}
For the third and fourth rule, one remembers that these terms always occur in an integral over $\vr$. With these rules the number of distinct terms reduces to three. We list them here, paired with with coefficients $k_{i}$ that identify them:
\begin{align}
&\kce \,\pda \ncia \pdb \ncib \nonumber \\
	& \quad = \kce (\mathrm{div} \vn)^{2} \label{eq:franksk1} \\[1ex]
&\kct \,\pda \ncib \pda \ncib = \kct (\pda \ncib)^{2} \nonumber \\
	& \quad = \kct \left((\mathrm{div} \vn)^{2} + (\vn \cdot \mathrm{curl} \vn)^{2} + (\vn \times \mathrm{curl} \vn)^{2} \right) \label{eq:franksk2} \\[1ex]
&\kcd \, \ncia \ncib \pda \ncig \pdb \ncig \nonumber \\
	& \quad = \kcd (\vn \times \mathrm{curl} \vn)^{2} \label{eq:franksk3}
\end{align}
Here we also wrote them in the usual vector notation, in which Frank and others defined the distortion free energy:
\begin{align}
f\td{d~(O-F)}(\vr) = \tfrac{1}{2} K_{1} (\mathrm{div} \vn)^{2} + \tfrac{1}{2} K_{2} (\vn \cdot \mathrm{curl} \vn)^{2} + \tfrac{1}{2} K_{3} (\vn \times \mathrm{curl} \vn)^{2}. \label{eq:frankfd-bis}
\end{align}
Using \eqref{eq:franksk1}--\eqref{eq:franksk3} in \refeq{eq:frankfd} from the main text, the Frank elastic constants $K_{i}$ can be expressed as:
\begin{align}
K_{1}&= 2(\kce + \kct) \label{eq:frankK1} \\
K_{2}&= 2 \kct  \label{eq:frankK4} \\
K_{3}&= 2(\kcd + \kct). \label{eq:frankK3}
\end{align}

\section{Listing all possible tensorial terms of the $\tM$-integrals}
\label{sec:listingdnterms}
In Sec.~\ref{sec:frankstate} we exemplified up to second rank a representation of the $\tM$-tensors --- of which the elements are the integrals of \refeq{eq:defMqelems} --- in terms of $J$-coefficients and a tensorial basis of $\vn(\vr)\otimes\vn(\vr)$ and $\mathbb{1}$, that is: composed of $n_{\mu}$'s and $\delta$'s. It is very useful to express the integral like this, because that allows for easy contraction with $\tH$ using the rules \eqref{eq:dnruleUnit}--\eqref{eq:dnruleEqUTsurf} introduced in the context of Oseen -- Frank theory. That this representation is possible is evident from the zeroth-order equation \eqref{eq:self0}, but also clear from its internal structure. As we show in Appendix~\ref{sec:appsymH}, the integral is zero once an odd number of any $\omcia$ is present; this yields diagonal ($\omcia \omcia$) elements. Expanding the exponent gives additional terms with factors $\vn \otimes \vn$ from $\tH$.

The notation of the $J$-coefficients is defined as $J_{N_{\delta},N_{m}}$, where $N_{\delta}$ is the number of indices associated to a Kronecker delta in the tensor, and $N_{m}$ counts the number of $m$'s. For higher ranks, a generalisation of \refeq{eq:MasJdn} is required. A similar exercise was done in Appendix A of \citeref{Garlea-Mulder_2017}, but here we require up to rank 8, as in \refeq{eq:order2}, or at least rank 6, as in \refeq{eq:order2iden}. It turns out that the possible combinations of $\delta$'s and $n_{\mu}$'s that construct a rank 8 tensor are so numerous that they no longer fit on one page. Rank 4 is still doable:
\begin{align}
M_{\alpha \beta \delta \gamma} =
\hspace{2.8ex}&J_{4,0} \left(\delta_{\alpha\beta} \delta_{\gamma\delta}+\delta_{\alpha\delta}
   \delta_{\beta\gamma}+\delta_{\alpha\gamma} \delta_{\beta\delta}\right) \label{eq:MwithJrank4} \\
+ &J_{2,2} \left(\delta_{\alpha\beta} n_{\delta}  n_{\gamma} +\delta_{\alpha\delta} n_{\beta}  n_{\gamma}
   +\delta_{\alpha\gamma} n_{\beta}  n_{\delta} +\delta_{\beta\delta} n_{\alpha}  n_{\gamma}
   +\delta_{\beta\gamma} n_{\alpha}  n_{\delta} +\delta_{\gamma\delta} n_{\alpha}  n_{\beta} \right) \nonumber \\
+&J_{0,4} \, n_{\alpha}  n_{\beta}  n_{\delta}  n_{\gamma}. \nonumber
\end{align}
In general, a tensor of rank $q$ that is composed in this way is a sum of all possible terms of the form
$$
\underbrace{\delta_{\mu_1 \mu_2} \dots \delta_{\mu_{\cdot} \mu_{\cdot}}}_{N_{\delta} \text{ indices}}  \underbrace{n_{\mu_{\cdot}} n_{\mu_{\cdot}} \dots n_{\mu_{q-1}} n_{\mu_q}}_{N_{m} \text{ indices}}
$$
for each choice of $N_{\delta}$ and $N_{n}$ such that both are even and $N_{\delta}+N_{n}=q$. This sum consists of groups characterised by one choice of $N_{\delta}$ and $N_{n}$, in which all elements have the \emph{same} coefficient $J_{N_{\delta},N_{n}}$, owing to the total symmetry in the indices of the $\tM$-tensor, cf. Appendix~\ref{sec:appsymH}. The number of terms in a group is\footnote{This relation is given in the appendix to the article by Han \emph{et al.} \cite{Han-ea_2014}, together with an exploration of the possible symmetric traceless tensors from $\delta$'s and pairs of unit vector elements.}

$$
\frac{(N_{n} + N_{\delta})!}{ \tfrac{N_{\delta}}{2}! \, N_{n}!\, 2^{N_{\delta}/2} }.
$$
For instance, there is a group with coefficient $J_{4,2}$ and $((2 + 4)!) \, / \, ( 2! \, 2!\, 2^2 ) = 45$ terms, of which $\delta_{\alpha\beta}\delta_{\gamma\delta} n_{\mu}  n_{\nu}$ is just one.

One can equate tensors like \eqref{eq:MwithJrank4} with the results of the next section, and solve for the $J_{N_{\delta},N_{n}}$-coefficients. This procedure is aided by a judicious diagonalisation of the fields, for instance by going to a basis from unit vectors that stand precisely in the $\mathbf{\hat{z}}$-direction, such that many elements on both sides of the equation are zero or similar, cf. Appendix~\ref{sec:intuniaxialH}. All the information about the $\tM$-integrals is contained in a list of these coefficients. Together with a straightforward algorithm \cite{Creyghton_2017m} that produces all possible tensors given for a given $N_{\delta}$ and $N_{n}$, they yield full and clear expressions for the $\tM$-tensors.

\section{Orientational integrals}
\label{sec:orientationalintegrals}
In the course of expanding terms in the expression for the free energy, integrals as in the elements \eqref{eq:defMqelems} of the $\tM$-tensor are encountered regularly. They are written as:
\begin{align}
 M_{\nu_{1} \dots \nu_{q}}(\vr) &\equiv  \int \ddom \, e^{- \tH(\vr) \, \colon \vomvom } \omega_{\nu_{1}} \dots \omega_{\nu_{q}} \text{,\,\,or}\nonumber \\
= \eM(\mathbf{q};\vr) &\equiv \int \ddom \, e^{- \tH(\vr) \, \colon \vomvom } \omce^{q_{1}} \omct^{q_{2}} \omcd^{q_{3}}.\label{eq:MMindtoq}
\end{align}
In expressions in the main text, $\tH(\vr)$ is mostly expanded such that in just $\tH\nnz(\vr)$ is present in these integrals. In the second line we introduced $\mathbf{q} = (q_1, q_2, q_3)$, a tuple of the number of $\omega$'s in each of the $3$ dimensions, which we counted (in a slight abuse of notation) as: $q_{i} \equiv \sum_{j=1}^{q} \delta_{\nu_{j} i} $ such that $\sum_{i=1}^{3} q_{i} = q$. In addition, through Eqs.~\ref{eq:defMqdHdHelems}\&\ref{eq:defMqddHelems} we introduced a notation for certain (contractions of) tensorial integrals involving vectors $\vom$ \emph{and} derivatives of fields $\tH$. The latter are independent of $\vom$ and can be taken out of the integral. In general, it is defined as
\begin{align}
&M^{[\lambda_{1},\ldots,\lambda_{l}]\;\ldots\;[\mu_{1},\ldots,\mu_{m}]}_{\nu_{1} \ldots \nu_{q}}  \label{eq:MdHsgeneral} \\
& \;\; \equiv 
  \Big(\partial^{l}_{\lambda_{1} \cdots \lambda_{l}} H_{\sigma_{1} \tau_{1}}\Big) 
  \ldots
  \Big(\partial^{m}_{\mu_{1} \cdots \mu_{m}} H_{\sigma_{n} \tau_{n}} \Big) 
 \int \ddom \, e^{-\tH \ctf{2} \vomvom} \,
 \omega_{\sigma_{1}}\omega_{\tau_{1}} \ldots \omega_{\sigma_{n}}\omega_{\tau_{n}} \,
 \omega_{\nu_{1}} \ldots \omega_{\nu_{q}}. \nonumber
\end{align}
We mixed tensor and index notation and omitted explicit $\vr$ dependence with the aim of readability. Note that the indices of the fields are all contracted with $\omega$'s in the integral. Here, (\emph{unlike} in Eqs.~\ref{eq:defMqdHdHelems}\&\ref{eq:defMqddHelems} in the main text) the directions of the derivatives (given by resp. $l, \ldots, m$ indices) may remain uncontracted, so the entire expression is a tensor of rank $l + \cdots + m + q$. However, equal indices can appear both up and down to $M$, in which case they are internally contracted dummies. Such is always the case in the main text, and there we omitted these dummies altogether to avoid clutter. Still, every $[\ldots]$ superscript implicitly adds two $\omega$'s to the integral, which is where the need for the higher rank expressions of Appendix~\ref{sec:listingdnterms} originates.

In this appendix, the $\tM$-integrals will be computed in a couple of cases. First, for a general symmetric $\mathbf{H}$ in 3 dimensions; this will result in a series expansion in the eigenvalues of $\mathbf{H}$. Then, in the special case where $\mathbf{H}$ is uniaxial and the axes can be rotated such that it has just one non-zero eigenvalue; this will give expressions in terms hypergeometric functions or error functions. The full results are given in a supplementary \textsl{Mathematica}-file \cite{Creyghton_2017m}.

Throughout the calculations we will use the following choice for axes and angular variables for the particle orientation $\vom$:
\begin{equation}
\vom =
\left(
\begin{array}{c}
 \sin \theta  \cos \varphi \\
 \sin \theta  \sin \varphi  \\
 \cos \theta  \\
\end{array}
\right) \label{eq:coordsvm}
\end{equation}
where $\theta \in [0,\pi]$ and $\varphi \in [0,2\pi)$ such that for instance $\vom(\theta=0)=\mathbf{\hat{z}}$ and the unit sphere reduces to the unit circle with azimuthal angle $\varphi$ for $\theta=\tfrac{\pi}{2}$.

\subsection{General considerations for a symmetric field tensor}
\label{sec:appsymH}
A symmetric rank-2 tensor $\tA$ is diagonalisable; we can denote it by $\tA_{\mathbb{d}}$ and its eigenvalues by $\alpha_{i}$ with $i = (1,2,3)$. Denoting the orientation vector as $\vom = \omci \vnci$ we can write the complete tensor of interest as:
\begin{align}
\mathbf{M}_{q}(\tA_{\mathbb{d}}) \ieqb \int \ddom \, e^{-\tA_{\mathbb{d}} \ctf{2} \vomvom} \bigotimes_{i=1}^{q} \vom \nonumber \\
 &\ieq \int \ddom \, e^{-\alpha_{i} \omci^{2}} \sum_{q_{1}+q_{2}+q_{3}=q} \omce^{q_{1}} \omct^{q_{2}} \omcd^{q_{3}} \bigotimes^{q_{1}} \vnce \bigotimes^{q_{2}} \vnct \bigotimes^{q_{3}} \vncd . \label{eq:defMkA}
\end{align}

Two remarks about the symmetries of this expression are in order. First, in the special case where all $\alpha_{i}=\alpha$ are equal, this tensor is fully symmetric in all coordinates, being an integration over the unit sphere. It suffices to calculate just one permutation of the $q_{i}$'s, and use the result for the symmetric other components of the resulting tensor as well. This symmetry can be exploited by writing a sum over permutations:
\begin{equation}
\mathbf{M}_{q}(\tA=\alpha \mathbb{1}) = \int \ddom \, e^{-\alpha_{i} \omci^{2}} \sum_{\substack{q_{1}+q_{2}+q_{3}=q \\ 0\leq q_{1} \leq q_{2} \leq q_{3} \leq q}}  \omce^{q_{1}} \omct^{q_{2}} \omcd^{q_{3}} \sum_{\substack{\text{permutations:} \\ \sigma(\{1,2,3\})}} \bigotimes^{q_{\sigma(1)}} \vnce \bigotimes^{q_{\sigma(2)}} \vnct \bigotimes^{q_{\sigma(3)}} \vncd , \label{eq:defMkAsym}
\end{equation}
thus reducing the distinct integrals that need to be computed. This simplification does not hold for unequal $\alpha_{i}$, but it partially resurfaces in cases where two of the eigenvalues are equal.

Second, the integrals are zero in all cases where at least one odd $q_{i}$ is present, as that implies an odd function in an even domain. Thus, many of the $\mathbf{M}$-tensor's elements are zero. This has two consequences. First, it is one of the conditions that allow for writing the tensor as a sum of products of vector elements and delta-functions, which was treated in Appendix \ref{sec:listingdnterms}. Second, it is essential to the rest of this section, in which we will be concerned with the \emph{elements} of tensors like these --- that is: with the values of the non-zero integrals, not with the tensorial nature of $\mathbf{M}$. The fact that all $q_{i}$ in these integrals are even allows for confining the domain of integration to just one octant of the sphere.

\subsection{Integration for a field tensor with three distinct non-zero eigenvalues.}
Now we focus on the computation of the value of the integral for a given combination of $q_{i}$'s. We denote $\vq=(\qce,\qct,\qcd)$ for this 3-tuple of exponents, and imply a similar meaning for the eigenvalues $\va$ and others quantities. For clarity, we will sometimes write them out explicitly, but more often we will use the brief notation.\\
For a component of the resulting tensor characterised by a given $\mathbf{q}$ and $\bss\alpha$ we can write:
\begin{align}
\mathcal{M}(\vq;\va) &= \int \ddom \, e^{-\alpha_{i} \omci^{2}} \omce^{\qce} \omct^{\qct} \omcd^{\qcd} \nonumber \\
&= \int_{0}^{2\pi} \dd \varphi \int_{0}^{\pi} \dd \theta \, \exp\Big[-\ace (\sin\theta \cos\varphi)^{2} -\act (\sin\theta \sin\varphi)^{2} -\acd  (\cos\theta)^{2} \Big] \nonumber \\
& \hspace{17.5ex} \times (\sin\theta )^{1 + \qce + \qct}  (\cos\theta)^{\qcd}  \, (\sin \varphi)^{\qct} (\cos \varphi)^{\qce} \label{eq:MexpAndCount}
\end{align} 
We worked out this general case using the work of Carlson \cite{Carlson_1972,Carlson}. Confining to one octant of the sphere, using that the $q_{i}$ are even, defining $\vp = \vq / 2$, and changing variables \emph{from} spherical coordinates \emph{to} linear coordinates over the 2-simplex, one finds following series expansion:
\begin{align}
&\mathcal{\tilde{M}}(\vp;\va)=  2 \pi \frac{\left(\sfrac{1}{2},p_{1}\right)\left(\sfrac{1}{2},p_{2}\right)\left(\sfrac{1}{2},p_{3}\right) }{\left(\sfrac{1}{2},\tfrac{n}{2}+1\right)} \hspace{25ex} \Big\}\equiv \gamma(\vp) \label{eq:ipa} \\
&\quad \times \sum_{l=0}^{\infty} \frac{(-1)^{l}}{\left({\textstyle \sum p_{i}} + \tfrac{3}{2},l\right)} \sum_{k_{1}+k_{2}+k_{3} = l} \frac{(p_{1}+\sfrac{1}{2}, k_{1})(p_{2}+\sfrac{1}{2}, k_{2})(p_{3}+\sfrac{1}{2}, k_{3})}{k_{1}!k_{2}!k_{3}!} \alpha_{1}^{k_{1}} \alpha_{2}^{k_{2}} \alpha_{3}^{k_{3}}.\nonumber
\end{align}
Just as most special functions, these Carlson function have no known analytical inverse, other than an inverse series expansion. Often a numerical root finding algorithm is the easiest way to obtain a result for the eigenvalues $\va$ of the field as a function of the required value for this integral.

In the above, we used \textsl{Appell's} symbol, a generalisation of the factorial defined as $(a,n) \equiv a(a+1)(a+2) \dots (a+n-1)$; the factorial is a special case for $a=1$: $(1,n)=n!$. By definition $(a,0)=1$. (N.B.: There are other names and notations for this object, such as \textsl{Pochhammer's} symbol and $(a)_{n}$.)

For later use, we note that $\mathrm{B}$ is the \textsl{Euler Beta} function, which is related to the $\Gamma$-function and Appell's symbol as:
\begin{align}
\mathrm{B}(b_{1}, \ldots, b_{n}) &\equiv \frac{\Gamma(b_{1}) \dots \Gamma(b_{n})}{\Gamma(b_{1} + \dots + b_{n})} \label{eq:betadef} \\
\mathrm{B}(p_{1}+\sfrac{1}{2},\,p_{2}+\sfrac{1}{2},\,p_{3}+\sfrac{1}{2}) &= \pi \frac{\left(\sfrac{1}{2},p_{1}\right)\left(\sfrac{1}{2},p_{2}\right)\left(\sfrac{1}{2},p_{3}\right) }{\left(\sfrac{1}{2},\sum p_{i}+1\right)} \nonumber.
\end{align}
In the case that all $p_{i}$ are zero the last line above equals $2\pi$.

\subsection{Integration for the Frank state}
\label{sec:intuniaxialH}
In the Frank state, we assume that the field $\tH$ can be written as \refeq{eq:HinFrank}. We wish to exploit the symmetry of $\tH$ to diagonalise it. This requires a rotation of the \emph{entire} problem, but this is of no consequence as in general the \qtens~theory is frame indifferent \emph{by construction}.\footnote{A nice review of this invariance and proofs thereof are given in \citeref{Mucci-Nicolodi_2012}.} The eigenvalues are $\tfrac{2}{3}U$ and twice  $- \tfrac{1}{3}U$. This degeneracy of eigenvalues in itself already somewhat simplifies the integrals tensor $\mathbf{M}$ by exploiting \emph{its} symmetry, as we will se shortly. A further algebraical simplification results from choosing the frame such that the first, distinct eigenvalue is placed on the `simplest' axis for $\vom$. That is: when we align to the vertical axis, the contraction with $\vomvom$ yields
\begin{equation}
\tH_{\mathbb{d}}(\vr) \ctf{2} \vomvom = U \left(\ncia \ncib - \tfrac{1}{3} \delta_{\alpha \beta}\right) \omcia \omcib = U(\mathbf{\hat{z}} \cdot \vom)^{2} - \frac{1}{3}U = U \cos^{2}(\theta) - \frac{U}{3}. \label{eq:HvomvomDiag}
\end{equation}
In the above subsections we already noted that the ordering of the eigenvalues is immaterial, but here we make the choice to let $\alpha_1 = \alpha_2 = - \tfrac{1}{3} U$ and $\alpha_3 = \tfrac{2}{3} U$. (The eigenvalues of $\mathbf{\hat{z}} \otimes \mathbf{\hat{z}}$ are $0$, $0$, and $1$, respectively.)

When \refeq{eq:HvomvomDiag} is used for the exponent in the $\mathcal{M}$, its last term can be taken out of the integral, giving $e^{U/3}$ as an overall factor.\footnote{This shift in $\va$ at the cost of introducing a factor $e^{U/3}$ is an example of the general property of Carlsons S-functions: $S(\vp,\va+\lambda)=e^{\lambda}S(\vp,\va)$, see Eq.~5.8-3 in \citeref{Carlson_1972}.} The result is a much simpler expression in which just $\alpha_3=U$ remains:
$$
\mathcal{\tilde{M}}(\vp;-\tfrac{1}{3}U,-\tfrac{1}{3}U,\tfrac{2}{3}U) = 
e^{U/3} \mathcal{\tilde{M}}(\vp;0,0,U) .
$$
Below, we show that this allows for writing them as better-known special functions such as the Kummer hypergeometric functions.\\

Working out \eqref{eq:ipa} in this simpler case we find
\begin{equation*}
\mathcal{\tilde{M}}(\vp;-\tfrac{1}{3}U,-\tfrac{1}{3}U,\tfrac{2}{3}U) = 
e^{U/3} \gamma(\vp) \sum_{l=0}^{\infty} \frac{(-1)^{l}}{l !} \frac{1}{\left({\textstyle \sum p_i } + \sfrac{3}{2},l\right)} (p_{3}+\sfrac{1}{2},l) \;  U^{l}.
\end{equation*}
Here it was used that the only surviving term in the sum over the $p_{i}$'s is that with $p_{3}=l$ and the other two zero; the others terms of this sum vanish via $\alpha_{1,2}=0$. Also, two factorials cancelled out, and $(a,0)=1$ was used. Note that $\sum p_{i} = \tfrac{q}{2}$ is still required, and that the expression is still dependent on $p_{1,2}$ via $\gamma(\vp)$ and $q$.

In the sum one can recognise\footnote{Cf. e.g. {http://functions.wolfram.com/HypergeometricFunctions/Hypergeometric1F1/06/01/02/01/01/0003/}.} the series expression of the hypergeometric function ${}_{1}F_{1}$, so that we can write
\begin{equation}
\mathcal{\tilde{M}}(\vp;-\tfrac{1}{3}U,-\tfrac{1}{3}U,\tfrac{2}{3}U) =  e^{U/3} \, \gamma(\vp) \; {}_{1}F_{1}\left(p_{3}+\frac{1}{2};{\textstyle \sum p_i } + \frac{3}{2}; -U \right)
 \label{eq:Mres3Duniax}
\end{equation}

Depending on $\vp$, the hypergeometric function can be rewritten as a collection of terms involving powers of $U$ and $\mathrm{erf}( \sqrt{U} )$. This follows from the definition of the error function as
\begin{equation}
\mathrm{erf}(z)= \frac{2 z}{\sqrt{\pi }} \sum _{l=0}^{\infty} \frac{(-1)^l}{l!} \frac{1}{2 l+1} z^{2 l},
\label{eq:deferror}
\end{equation}
which is easily matched to the expression for $\mathcal{\tilde{M}}(\mathbf{0}; 0,0, U)$, and, with a bit more algebra, also to cases with nonzero $\vp$. In practice, \textsl{Mathematica} or \textsl{Wolfram}'s reference tables are employed for this exercise.

\subsubsection{Alternative, direct derivation}
Running the risk of exhausting the reader we will offer yet another way of obtaining \eqref{eq:Mres3Duniax} that does not need the Carlson's functions, which are both very general and arguably complex. The direct derivation starts with the observation that there are integral representations\footnote{See {http://functions.wolfram.com/GammaBetaErf/Beta/07/01/01/}\\ and {http://functions.wolfram.com/HypergeometricFunctions/Hypergeometric1F1/07/01/01/}.} of the Beta and hypergeometric functions that are reminiscent of the type of integrals we want to solve:
\begin{align}
\mathrm{B}(a,b)
&= \int_{0}^{1} t^{a-1} (1-t)^{b-1} \, \dd t \label{eq:betaintt} \\
&= 2 \int _{0}^{\sfrac{\pi }{2}} (\sin\theta )^{2a-1} (\cos\theta )^{2b-1}\,\dd t \label{eq:betasincos} \\
\text{and} \nonumber\\
{}_{1}F_{1}(a;b;z)
&= \frac{\Gamma(b)}{\Gamma(a)\Gamma(b-a)} \int_{0}^{1} e^{z t}\, t^{a-1} (1-t)^{-a+b-1}\, \dd t . \label{eq:kummerconfint}
\end{align}
Representation \eqref{eq:betasincos} is directly applicable to our integral in the case of zero $\mathbf{H}$ for both the polar and azimuthal angles. Once more using the symmetry of all 8 octants of the sphere and the evenness of the integrand we write:
\begin{align*}
 \int _{0}^{\pi} (\sin\theta )^{\eta_{x}} (\cos\theta )^{\eta_{y}}\,\dd\theta &  \int_{0}^{2 \pi} (\sin\varphi )^{\upsilon_{x}} (\cos\varphi )^{\upsilon_{y}}\,\dd\varphi \\
 = \mathrm{B}\left(\tfrac{\eta_{x}+1}{2},\tfrac{\eta_{y}+1}{2}\right) & \, 2 \, \mathrm{B}\left(\tfrac{\upsilon_{x}+1}{2},\tfrac{\upsilon_{y}+1}{2}\right),
\end{align*}
and counting the powers of the desired sines and cosines in our integral we arrive at
\begin{equation*}
\mathcal{M}(\vq;\mathbf{0}) =  2 \,\mathrm{B}\left(\frac{q_1 + q_2}{2}  + 1 ,\, \frac{q_3 + 1}{2}\right) \, \mathrm{B}\left(\frac{q_1 + 1}{2} , \, \frac{q_2 + 1}{2} \right).
\end{equation*}
The Beta function is symmetric under interchange of its two arguments --- a reflection of the symmetry we already observed in \refeq{eq:defMkAsym}.

For nonzero $\alpha_{3}$ we need the integral representation ${}_{1}F_{1}$, at least for the polar angle $\theta$. We rewrite its integral; a step analogous to \eqref{eq:betaintt} from \eqref{eq:betasincos}:

\allowdisplaybreaks
\begin{align*}
&\hphantom{+2} \int_{0}^{\pi} \sin\theta \dd\theta \; e^{- U \cos^{2}\theta} \; (\sin\theta)^{q_1 + q_2} (\cos\theta)^{q_3} \\
=& -2 \int_{1}^{0} \dd\xi \;e^{- U \xi^{2}} \; (\sqrt{1-\xi^{2}})^{q_1 + q_2}\;  \xi^{q_3} \\
=& +1 \int _{0}^{1} \dd t \; e^{- U t} \; (1-t)^{\tfrac{q_1 + q_2}{2}} \; t^{\tfrac{q_3-1}{2}}\\
=&\; \frac{\Gamma\left(\tfrac{q_3+1}{2}\right)\Gamma\left(\tfrac{q_1 + q_2+2}{2}\right)}{\Gamma\left(\tfrac{q_1 + q_2+q_3+3}{2}\right)} \; {}_{1}F_{1}\left(\frac{q_3+1}{2}; \frac{q_1 + q_2+q_3+3}{2}; -U \right).
\end{align*}
We used (i) in the first equality, first a restriction to the first half of the integration domain, and then a change of variables $\xi = \cos\theta$ such that the Jacobian changed as $\sin\theta \dd\theta = - \dd \xi$ and the domain changed to $\cos(0)=1$ through $\cos(\sfrac{\pi}{2}) = 0$; (ii) in the second equality, a change of variables $t=\xi^{2}$ such that $\dd \xi = \tfrac{1}{2 \sqrt{t}} \dd t$; and (iii) in the last equality, an application of definition \eqref{eq:kummerconfint}.
We found the same arguments for the hypergeometric function as in \eqref{eq:Mres3Duniax}, namely $a=\tfrac{q_{3}+1}{2}=p_{3} + \sfrac{1}{2}$ and $b=\tfrac{q_{1}+q_{2}+q_{3}+3}{2}={\textstyle \sum p_i } + \tfrac{3}{2}$. Moreover, once the Beta function for the azimuthal angle is included, the factor in front of the function agree with \eqref{eq:Mres3Duniax} as well: using the definition \eqref{eq:betadef} one can easily show that indeed
\begin{equation}
\gamma(\vp) = 2 \pi  \frac{ \left(\tfrac{1}{2}, p_{1} \right) \left(\tfrac{1}{2}, p_{2} \right) \left(\tfrac{1}{2}, p_{3} \right)  }  
   {\left(\tfrac{1}{2}, p_{1} +p_{2} +p_{3} +1\right) }
   = 2 \, \mathrm{B}\left(p_1 + \tfrac{1}{2} , \, p_2 + \tfrac{1}{2} \right) \, \frac{\Gamma\left(p_3 +\tfrac{1}{2}\right)\Gamma\left(p_1 + p_2 + 1 \right)}{\Gamma\left(p_1 + p_2 + p_3 + \tfrac{3}{2} \right)}. \label{eq:gammap}
\end{equation}

\section{Full results for the second-order equation}
\label{sec:fullsolutions}
Here we give the full solutions to the second-order equation \eqref{eq:order2}, assuming the form $\tH^{(2)}(\vr) = V \, \vn(\vr)\otimes\vn(\vr)+W \,\mathbb{1}$. Both the trace of the equation (cf. \refeq{eq:order2iden})  and the contraction using $\ncis\ncit$ need to hold; together they fix both $V$ and $W$.
\begin{align}
V=\hspace{1.8ex}
& \frac{8 U^2 S^2-60 U S^2-4 U^2 S+24 U S+135 S-4 U^2+18 U}{96 \left(4 U S^2-2 U S-9 S-2 U\right)} 
    \partial_{\alpha} n_{\alpha}  \partial_{\beta} n_{\beta} \label{eq:Vsol} \\
+& \frac{8 U^2 S^2-60 U S^2-4 U^2 S+24 U S+135 S-4 U^2+18 U}{192 \left(4 U S^2-2 U S-9 S-2 U\right)}
    (\partial_{\alpha} n_{\beta})^2 \nonumber \\
+& \frac{56 U^2 S^2+300 U S^2-28 U^2 S-192 U S-675 S-28 U^2-90 U}{192 \left(4 U S^2-2 U S-9 S-2 U\right)}
    n_{\alpha}n_{\beta} \partial_{\alpha} n_{\gamma}  \partial_{\beta} n_{\gamma} \nonumber \\
W=
-&\frac{8 U^2 S^2-12 U S^2-4 U^2 S-12 U S+45 S-4 U^2+6 U}{96 \left(4 U S^2-2 U S-9 S-2 U\right)}
    \partial_{\alpha} n_{\alpha}  \partial_{\beta} n_{\beta} \label{eq:Wsol} \\
-& \frac{8 U^2 S^2-12 U S^2-4 U^2 S-12 U S+45 S-4 U^2+6 U}{192 \left(4 U S^2-2 U S-9 S-2 U\right)}
    (\partial_{\alpha} n_{\beta})^2 \nonumber \\
+& \frac{8 U^2 S^2-156 U S^2-4 U^2 S-36 U S+225 S-4 U^2+30 U}{192 \left(4 U S^2-2 U S-9 S-2 U\right)}
    n_{\alpha}n_{\beta} \partial_{\alpha} n_{\gamma}  \partial_{\beta} n_{\gamma} \nonumber
\end{align}
The solutions for $V$ and $W$ are initially expressed in terms of the $J$-coefficients, cf. Appendix~\ref{sec:listingdnterms}. The various terms are grouped in three according to Appendix~\ref{sec:frankrules}; this importantly contributes to the numerical factors in the above result. The values for the coefficients stem from the result of Appendix~\ref{sec:orientationalintegrals}. Lastly, for aesthetical reasons all exponents and special functions are effectively hidden from view by identifying factors $S$, the zeroth-order result \refeq{eq:SfromUzero}.

\section{Generalisation to discs and other symmetric particles}
\label{sec:discs}
We are considering inversion symmetric and cylindrically symmetric particles. Their orientation can be specified by a single unit vector $\vom$ along the symmetry axis. The remaining axes of the particles can be specified by choosing a single unit vector $\vxi$ orthogonal to $\vom$, which then fixes a third molecular frame vector
$\veta=\vom \wedge \vxi$. 
Naturally, $\vxivxi + \vetaveta + \vomvom = \mathbb{1}$ in any frame. The particle is thus generically defined by the set
\begin{equation}
\prtset=\left\{  \prt= u \cos\varphi \, \vxi + u \sin \varphi \, \veta + w \, \vom \; \big| \, \left(u, \varphi, w \right)
\in\Omega\subset\mathbb{R}^{3} \right\}
\end{equation}
For a cylindrical particle of radius $R$ and length $L$, $\Omega=[0, R] \times [0, 2\pi] \times [-\sfrac{1}{2} L, \sfrac{1}{2} L]$. Inversion symmetry implies that if $\prt \in \prtset$ then also $-\prt \in \prtset$. We also introduce the measure of the particle
\begin{equation}
\mu\left( \prtset \right)  = \int_{\Omega} \ddprtdeps.
\end{equation}
We introduce the following notation as a generalisation of \refeq{eq:LengthAvNot}:
\begin{equation}
\big[ f \left( \prt, X \right) \big]_{\prtset}
=\frac{1}{\mu \left( \prtset \right)}
\int_{\Omega}\ddprtdeps
f \left( \prt, X \right)
\end{equation}
We now note that by assumption of the inversion symmetry alone
\begin{equation}
\big[\prt \big]_{\prtset}=0
\end{equation}
The only relevant quantity we need to evaluate in our expansions to second
order in the dimensions of the particle therefore is
\begin{align}
& \big[ \prtprt \big]_{\prtset} \nonumber\\
&=\frac{1}{\mu\left(
\prtset\right)} \int_{\Omega} \ddprtdeps
\left\{ \begin{array}[c]{c}
u^{2}\cos^{2}\varphi \, \vxivxi + u^{2}\sin^{2}\varphi \, \vetaveta + w^{2} \, \vomvom \\
+2u^{2}\cos\varphi\sin\varphi \, \vxi\otimes\veta + 2 uw \cos\varphi \, \vxi\otimes\vom + 2 uw \sin\varphi \, \veta \otimes \vom
\end{array} \right\}  \nonumber\\
& = M_{\bot}\left(\mathbb{1} - \vomvom\right) + M_{\Vert} \, \vomvom \label{eq:genppp}
\end{align}
where $M_{\bot}$ and $M_{\Vert}$ are the contributions from the perpendicular and parallel directions, respectively. The three terms on the second line between the curly brackets give zero from the integration over $\varphi$, hence the reduction to the two terms with $\vomvom$ and $\vxivxi + \vetaveta = \mathbb{1} - \vomvom$.

For a cylinder $\Omega=[0, R] \times [0, 2\pi] \times [-\sfrac{1}{2} L, \sfrac{1}{2} L]$ we have
$$\mu\left(\prtset \right)  =\int_{0}^{R} u \, \dd u \int_{0}^{2\pi} \dd \varphi \int_{-\sfrac{1}{2}L}^{+\sfrac{1}{2}L} \dd w= \pi R^{2} L,$$
and the general second moment result \eqref{eq:genppp} becomes
\begin{equation}
\big[ \prtprt \big]_{\prtset} = \frac{1}{4} R^{2} \left(\mathbb{1} - \vomvom\right) + \frac{1}{12} \, L^{2}\vomvom.
\label{eq:cylppp}
\end{equation}
For flat cylindrical discs ($\Omega=[0, R] \times [0, 2\pi] \times \{0\}$, so $L=0$ and finite $R$) the above expression reduces to the first term. For the rigid rods that we treated in the main text ($R=0$ and finite $L$ instead, so $\Omega=\{0\} \times \{0\} \times [-\sfrac{1}{2} L, \sfrac{1}{2} L]$) this is just the second term. This corresponds with the expressions of e.g. Sec.~\ref{sec:perturbativegeneral}, up to a Taylor-factor $\sfrac{1}{2}$. Note that \eqref{eq:cylppp} is similar, but not equal, to the moment of inertia tensor for a solid cylinder.

%
%
\subsection{Application in free energy expansion}
Generalising the main text of this work from rigid rods to other rigid symmetrical (cylindric, head-tail) particles, we start from the $L$-density \refeq{eq:spdfwithlength}, which now becomes a $\prtset$-density
\begin{align}
o_{\prtset}(\vr,\vom) \equiv \rho \big[  \psi(\vr - \prt,\vom) \big]_{\prtset}. \label{eq:spdfwithprt}
\end{align}
The derivations of Sec.~\ref{sec:nonlocal} are ported immediately, yielding this generalisation of \refeq{eq:Ostartdirect}:
\begin{align}
 \tO(\vr) = N \frac{\int \ddom \, \left[ e^{-\beta [\tH(\vr + \prt\pr \hspace{-0.25ex}- \prt)]_{\prtset} \ctf{2} \vomvom} \right]_{\prtset\pr} \, \vomvom}{\int \ddr\pr \int \ddom \, e^{-\beta [\tH(\vr\pr - \prt )]_{\prtset} \, \colon \vomvom}},
 \label{eq:Odirectppp}
\end{align}

To write the generalisation of the direct expansion as in Eqs.~\ref{eq:numexpand}--\ref{eq:defD}, we need to employ the more general definition of the $M$-tensors given in \refeq{eq:MdHsgeneral}, in which the derivatives' indices are explicitly written and may be either contracted with $\omega$'s \emph{in} the $M$-tensor, or with something \emph{else}, e.g. a $\delta$; it is the latter we need for the $\mathbb{1}$-term from $\left[ \prtprt \right]_{\prtset}$. For the numerator resp. denominator of \refeq{eq:Odirectppp} we now work out the expansion and find up to second-order
\begin{align}
 \int \ddom \, \left[ e^{-\beta [\tH(\vr + \prt\pr \hspace{-0.25ex}- \prt)]_{\prtset} \ctf{2} \vomvom} \right]_{\prtset\pr} \, \omcis \omcit = \hspace{20ex}& \label{eq:expdirectnumppp} \\
 M_{\sigma\tau}(\vr) + \frac{1}{24} \Big( \left( L^2 - 3 R^2 \right) \left\{ M_{\alpha\beta\sigma \tau}^{[\alpha][\beta]}(\vr) - 2 M_{\alpha\beta\sigma \tau}^{[\alpha\beta]}(\vr)  \right\}& \nonumber \\
 + 3 R^2 \delta_{\alpha\beta}  \left\{ M_{\sigma \tau}^{[\alpha][\beta]}(\vr) - 2 M_{\sigma \tau}^{[\alpha\beta]}(\vr)\right\}& \Big), \nonumber \\
\int \ddr\pr \int \ddom \, e^{-\beta [\tH(\vr\pr - \prt )]_{\prtset} \, \colon \vomvom} = \hspace{32,5ex} & \label{eq:expdirectdenomppp}\\
\int \ddr\pr M(\vr) - \frac{1}{24} \Big( \left( L^2 - 3 R^2 \right) M_{\alpha\beta}^{[\alpha\beta]}(\vr\pr) + 3 R^2 \delta_{\alpha\beta} M^{[\alpha\beta]}(\vr\pr)& \Big),\nonumber
\end{align}
which indeed agrees with the structure offered by result \ref{eq:cylppp}. Note that the reasoning at identities \ref{eq:MtoSurfaceIdentity}\&\ref{eq:order2iden} holds, so Eqs.~\ref{eq:expdirectnumppp}\&\ref{eq:expdirectdenomppp} are just two manifestations of one requirement. Also note that there is no surviving term that couples $R$ to $L$. The first such terms would be proportional to $L^2 R^2$; further expansion would reveal many of those, but being of fourth order they are discarded here. Hence, to second order, the perturbative solution strategy using
\begin{equation}
\tH(\vr) = \tH\nnz(\vr) + L^2 \tH\nn{2,L}(\vr) + R^2  \tH\nn{2,R}(\vr) + \boh(L^{n}R^{m} \, | \, m+n>2).
\label{eq:perturbLandRexp}
\end{equation}
will simply yield a set of $R^{2}$-dependent elastic terms in addition to the $L^{2}$-terms we already found in the main text. Next to requirement \refeq{eq:order2}, we now have for $R^{2}$:
\begin{align}
-H_{\alpha \beta}\nn{2,R}(\vr) M_{\alpha \beta \sigma \tau}(\vr) + \frac{1}{8} \bigg( \delta_{\alpha\beta}  \left\{ M_{\sigma \tau}^{[\alpha][\beta]}(\vr) - 2 M_{\sigma \tau}^{[\alpha\beta]}(\vr)\right\}& \label{eq:order2Req} \\
 -  \left\{ M_{\alpha\beta\sigma \tau}^{[\alpha][\beta]}(\vr) - 2 M_{\alpha\beta\sigma \tau}^{[\alpha\beta]}(\vr)  \right\}& \bigg) = 0. \nonumber
\end{align}
Note that the second line is simply $-3 H_{\alpha \beta}\nn{2,L}(\vr)$; see \refeq{eq:sol2} for the result. The derivative's indices of the $M$-terms on the first line are contracted by a $\delta$, as a result of which they only contribute to the $k_{2}$-term, proportional to $(\partial_{\alpha} n_{\beta})^2$. This leads to a strong contribution to the $K_{2}$ elastic constant, cf. Fig.~\ref{fig:constantsplotdiscs}, whereas the other two constants are more or less suppressed by the terms on the second line.

\end{document}